\documentclass[numsec,webpdf,modern,large]{article}

\usepackage{multirow}
\usepackage{authblk}
\usepackage{graphicx}
\usepackage{amsmath}
\usepackage{amsfonts}
\usepackage{xcolor}
\usepackage{float}
\usepackage{color,soul}
\usepackage{hyperref}
\usepackage{comment}
\usepackage{csvsimple}
\usepackage{caption}
\usepackage{subcaption}
\usepackage{stmaryrd}
\usepackage[export]{adjustbox}
\usepackage[margin=1in]{geometry}
\usepackage{amsmath}
\usepackage{bbm}
\usepackage{natbib}
\usepackage[customcolors]{hf-tikz}
\usepackage[linesnumbered,ruled,vlined]{algorithm2e}
\usepackage{todonotes}
\usepackage{url}
\usepackage{adjustbox}

\hfsetfillcolor{green!10}
\def\BE{\pmb{E}}

\newcounter{suppfigure}

\title{Approximate Bayesian Computation sequential Monte Carlo via random forests}
\author[1]{Khanh N. Dinh}
\author[1,2]{Cécile Liu}
\author[1,3]{Zijin Xiang}
\author[1,4]{Zhihan Liu}
\author[1]{Simon Tavaré}
\affil[1]{Irving Institute for Cancer Dynamics and Department of Statistics, Columbia University, New York, NY, USA}
\affil[2]{Department of Applied Mathematics, Ecole Nationale des Ponts et Chaussées, Champs-sur-Marne, France}
\affil[3]{Department of Quantitative and Computational Biology, University of Southern California, CA, USA}
\affil[4]{Department of Applied Mathematics and Statistics, Stony Brook University, NY, USA}
\date{\today}

\begin{document}

\maketitle

\newpage

\tableofcontents

\newpage

\abstract{
Approximate Bayesian Computation (ABC) is a popular inference method when likelihoods are hard to come by.
Practical bottlenecks of ABC applications include selecting statistics that summarize the data without losing too much information or introducing uncertainty, and choosing distance functions and tolerance thresholds that balance accuracy and computational efficiency.
Recent studies have shown that ABC methods using random forest (RF) methodology perform well while circumventing many of ABC's drawbacks.
However, RF construction is computationally expensive for large numbers of trees and model simulations, and there can be high uncertainty in the posterior if the prior distribution is uninformative.
Here we further adapt random forests to the ABC setting in two ways. The first exploits distributional random forests to provide a direct method for inferring the joint posterior distribution of parameters of interest, while the second describes a sequential Monte Carlo approach which updates the prior distribution iteratively to focus on the most likely regions in the parameter space.
We show that the new methods can accurately infer posterior distributions for a wide range of deterministic and stochastic models in different scientific areas.
}

\section{Introduction}
Mathematical modeling has played an important role in studying scientific phenomena.
Its practical applications often depend on accurately extracting model parameters $\theta$ from experimentally observed data, $y^{\rm obs}$.
In the Bayesian framework, this entails inferring the posterior distribution 
$$
\pi(\theta \mid y^{\rm obs}) \propto f(y^{\rm obs} \mid \theta) \cdot \pi(\theta)
$$ 
of the parameters from the data, exploiting the likelihood $f(y^{\rm obs} \mid \theta)$ and the prior $\pi(\theta)$ of such observations under the model with given parameter values.
However, the likelihood function is often intractable to derive theoretically, difficult to compute numerically, or too complex to optimize directly.
Approximate Bayesian Computation (ABC) was proposed as an alternative method to approximate the posterior distributions in such scenarios \citep{tavare1997inferring,FuLi1997,pritchard1999population,Beaumont2002}.
Using statistics to summarize model simulations, ABC seeks the parameters that result in minimal distance between the statistics and those of the observed data.
Different variations have been proposed to improve the performance of the original ABC method, such as ABC Monte Carlo Markov Chain \citep{marjoram2003markov} and ABC Sequential Monte Carlo \citep{toni_approximate_2008}, and they have collectively found remarkable success in applications across different scientific areas \citep{sisson2018handbook}.

The accuracy and efficiency of ABC in practice depends on several factors.
First, one requires a distance function to compare observed and simulated data.
The metric is typically weighted so that different statistics contribute equally \citep{prangle2017adapting}.
However, an equally important consideration is the relevance of each statistic, defined as the amount of information that it carries toward identifying the underlying parameters.
Optimizing these two criteria can be challenging \citep{jung2011choice}.
Second, ABC requires a tolerance threshold to decide whether proposed parameters are accepted or rejected.
The tolerance $\epsilon$ poses a trade-off between the computational efficiency, which increases with higher $\epsilon$, and accuracy of the empirical posterior distributions, which improves as $\epsilon$ decreases.
Choosing $\epsilon$ depends on the specific model, statistic choices and distance function, and it may require intuition or experimentation \citep{jung2011choice}.
Finally, and most importantly, ABC's results depend on the choice of summary statistics.
The approximated posterior distribution is only guaranteed to converge to the likelihood-based distribution as $\epsilon \to 0$ if the statistics are sufficient.
For complex models, it is not always possible or realistic to find low-dimensional sufficient statistics 
\citep[Chapter 5]{sisson2018handbook}.
This can sometimes be remedied by inclusion of many distinct statistics.
However, the posterior estimation may become distorted if the selected statistics are noisy or uninformative.
These factors in combination mean that optimizing the performance of ABC in practice can be challenging and require extensive experimentation.

Recent studies have shown that random forests (RF), a powerful non-parametric regression method \citep{breiman2001random}, can be employed in the ABC context to infer posterior distributions \citep{raynal2019abc}. 
Its appeal lies in reduced dependence  on user-defined hyperparameters that are essential to traditional ABC implementations, including the metric function and tolerance threshold.
Moreover, RF has been shown to perform well even if the majority of statistics are pure noise, indicating that it is significantly more tolerant to a wider selection of statistics to represent the data \citep{raynal2019abc}.

Therefore, RF methods have the potential to approximate the posterior distributions well with lower computational cost.
However, the inclusion of all simulations in inferring parameters, coupled with wide range in the prior distributions for some models, can sometimes result in higher uncertainty in the posterior distributions.

In this paper, we propose a new  inference method, Approximate Bayesian Computation sequential Monte Carlo with random forests (ABC-SMC-(D)RF).
It inherits the non-parametric nature of RF methods, but is embedded in the framework of Sequential Monte Carlo \citep{toni_approximate_2008}.
The posterior distribution evolves with successive iterations to focus on the most likely regions in the parameter space, resulting in more relevant model simulations and lower uncertainty in the final distributions.

We use some traditional ABC methods in comparisons with ABC-SMC-(D)RF described in the following sections.
Recent ABC implementations using random forests are reviewed in Section \ref{sec:rf}, with illustrative examples in Section \ref{sec:abcdrf}.
In Section \ref{sec:abcsmcrf}, we describe ABC-SMC-(D)RF's methodology.
We then demonstrate its performance, compared both to traditional ABC algorithms and previous RF-based methods, in a variety of deterministic and stochastic models (Section \ref{sec:results}).
We conclude with a discussion.

\section{Random forests in the context of ABC}\label{sec:rf}

Although simple to implement, the ABC methods depend heavily on the choices of hyperparameters, including the summary statistic function $S(y)$, the distance function $d(s,s')$ and the tolerance threshold $\epsilon$ \citep{sisson2018handbook}.
Furthermore, the choices of Markov kernels to propose new particles in ABC Monte Carlo Markov Chain (ABC-MCMC) and ABC Sequential Monte Carlo (ABC-SMC) also affect the approximated posterior distributions.
As random forests (RF) gain in popularity as a powerful non-parametric regression technique \citep{breiman2001random, segal2004machine, rigatti2017random, desai2021regional}, its applications in the context of ABC are also becoming more prominent.
RF-based methods eliminate the need for choice of metric, tolerance level and the perturbation kernels, making them an attractive alternative to the previous ABC implementations.
Importantly, the RF method is relatively robust to noise \citep{marin2018likelihood}, in that $S(y)$ can include many poorly informative statistics without significant impact on the results. 

\subsection{One-dimensional ABC random forests}

Raynal et al. introduced ABC random forest (ABC-RF) to approximate the posterior distribution for ABC inference problems with one parameter \citep{raynal2019abc}, based on the random forest formulation (Algorithm \ref{alg:abc-rf-1}) developed by \cite{breiman2001random}.
The root node for each tree $T_t$ is the reference table, consisting of parameters drawn from the prior distribution and corresponding statistics.
Each node $R_*$ in the decision tree is split into two nodes $R_{*1}$ and $R_{*2}$ by selecting one statistic $k_*\in \{1,\dots,|S|\}$, where $|S|$ is the total number of summary statistics, 
and a threshold $s_*$ such that the $L^2$-loss between the divided sets
\begin{equation}\label{eq:L2}
    \frac{1}{|R_*|}\left(\sum_{(\theta,s)\in R_{*1}}\left(\theta-\bar\theta_1\right)^2 +
    \sum_{(\theta,s)\in R_{*2}}\left(\theta-\bar\theta_2\right)^2\right)
\end{equation}
is minimized. Here $|R_*|$ is the number of particles in $R_*$, and $\bar\theta_1$, $\bar\theta_2$ are the parameter means in $R_{*1}$ and $R_{*2}$ respectively.
The tree is finished when each leaf either consists of less than $N_{\rm min}$ particles, or all of the leaf's particles have the same statistics.
To guarantee forest diversity, the root node of each tree (Step 5 in Algorithm \ref{alg:abc-rf-1}) is bootstrapped from the reference table, and the statistics considered for splitting each node
(Step 9 in Algorithm \ref{alg:abc-rf-1}) is limited to a randomly selected subset of size $n_{\rm try}$ among the available statistics $S$.

\bigskip
\begin{minipage}{0.8\textwidth}
\begin{algorithm}[H]
    \caption{Growing regression trees for ABC-RF \citep{raynal2019abc}}\label{alg:abc-rf-1}
    Sample $\theta^{(1)},\dots,\theta^{(N)}\sim\pi(\theta)$
    
    Simulate $y^{(1)},\dots,y^{(N)}$ and compute $s^{(1)},\dots,s^{(N)}$
    
    Form reference table $R=\left\{\left(\theta^{(1)},s^{(1)}\right),\dots,\left(\theta^{(N)},s^{(N)}\right)\right\}$
    
    \For{$t=1,\dots,B$}{
        $R_0\gets$ bootstrapped from $R$
    
        Tree $T_t$ is rooted in node $R_0$
        
        \While{$\max_{\mathrm{leaf }R_*\in T_t}|R_*|>N_{\rm min}$}{
            Find a leaf $R_*$ in $T_t$ with $\left|R_*\right|>N_{\rm min}$

            Statistic candidates for splitting $R_*\gets$ sample of size $n_{\rm try}$ among $S$

            Choose statistic $k_*$ among the candidates and threshold $s_*\in\mathbb{R}$ to split node $R_*$ into new leaves $R_{*1}=\left\{(\theta,s)\in R_*:s_{k_*}\le s_*\right\}$ and $R_{*2}=\left\{(\theta,s)\in R_*:s_{k_*}> s_*\right\}$, such that the $L^2$-loss criterion (Eq. \eqref{eq:L2}) is minimized
        }
    }
\end{algorithm}
\end{minipage}
\bigskip

Given an observation $y^{\rm obs}$, predicting its parameter $\theta$ based on tree $T_t$ involves following from the root node with the statistics $s^{\rm obs}$ and comparing $s^{\rm obs}_{k_*}$ with $s_*$ at each node $R_*$ to select the next node, until reaching a leaf $L_t(s^{\rm obs})$.
The prediction for $\theta$ based on $T_t$ is then the average parameter among particles in $L_t(s^{\rm obs})$:
\begin{equation*}
    \frac{\sum_{i=1}^N\theta^{(i)}\cdot n_t^{(i)}\cdot\textbf{1}_{s^{(i)}\in L_t(s^{\rm obs})}}{\sum_{i=1}^N n_t^{(i)}\cdot\textbf{1}_{s^{(i)}\in L_t(s^{\rm obs})}}
\end{equation*}
where $n_t^{(i)}$ is the number of times $\theta^{(i)}$ is duplicated in the bootstrapped sample $R_0$, and $\textbf{1}_{s^{(i)}\in L_t(s^{\rm obs})}$ is the indicator for whether $s^{(i)}$ falls into the same leaf as $s^{\rm obs}$.
The prediction for $\theta$ based on the whole forest is the average of predictions based on each tree:
\begin{equation*}
    \sum_{i=1}^N\theta^{(i)}\cdot
    \frac{1}{B}
    \sum_{t=1}^{B}
    \frac{n_t^{(i)}\cdot\textbf{1}_{s^{(i)}\in L_t(s^{\rm obs})}}{\sum_{i=1}^N n_t^{(i)}\cdot\textbf{1}_{s^{(i)}\in L_t(s^{\rm obs})}}
\end{equation*}
\cite{raynal2019abc} argued that this weighted estimate implies that the density of particles $\theta^{(i)}$ with corresponding weights $w^{(i)}$ (Algorithm \ref{alg:abc-rf-2}) forms the approximation $\pi_{ABC-RF}\left(\theta\middle|s^{\rm obs}\right)$ for the posterior distribution.

\bigskip
\begin{minipage}{0.8\textwidth}
\begin{algorithm}[H]
    \caption{Posterior distribution from ABC-RF \citep{raynal2019abc}}\label{alg:abc-rf-2}
    \For{$t=1,\dots,B$}{
        Follow the tree $T_t$ from root node with $s^{\rm obs}$ until locating leaf $L_t(s^{\rm obs})$
    }
    \For{$i=1,\dots,N$}{
        Weight for particle $\theta^{(i)}\gets w^{(i)}=\frac{1}{B}\sum_{t=1}^{B}\frac{n_t^{(i)}\cdot\textbf{1}_{s^{(i)}\in L_t(s^{\rm obs})}}{\sum_{i=1}^N n_t^{(i)}\cdot\textbf{1}_{s^{(i)}\in L_t(s^{\rm obs})}}$
    }
\end{algorithm}
\end{minipage}
\bigskip

\cite{abcrf-packages}  developed an {\sf R} package \texttt{abcrf} based on the ABC-RF method.
For multivariate problems, ABC-RF is usually applied for each parameter marginally \citep{raynal2019abc}.
The authors also provided suggested values for RF hyperparameters;
by default, $N_{\rm min}=5$ and $n_{\rm try}=|S|/3$.
Among the most important hyperparameters is the count of trees $B$, for which they recommend analyzing whether the out-of-bag mean squared error stabilizes around the selected $B$ \citep{raynal2019abc,pudlo2016reliable}.

\subsection{Distributional random forests}
\cite{cevid_distributional_2022} developed distributional random forests (DRF) for multivariate regression problems. The approach considers two criteria to split tree nodes. 
The first criterion extends the $L^2$-loss formula (Eq. \ref{eq:L2}), which the authors rewrite as 
\begin{equation*}
    \frac{1}{|R_*|}\sum_{(\theta,s)\in R_{*}}\left(\theta-\bar\theta\right)^2 -
    \frac{|R_{*1}|\cdot|R_{*2}|}{|R_{*}|^2}\left(\bar\theta_2-\bar\theta_1\right)^2
\end{equation*}
where $\bar\theta$ is the parameter mean in the parent node $R_*$.
Because the first term does not depend on the split, minimizing the $L^2$-loss is equivalent to maximizing
\begin{equation}\label{eq:CART}
    \frac{|R_{*1}|\cdot|R_{*2}|}{|R_{*}|^2}\left(\bar\theta_2-\bar\theta_1\right)^2
\end{equation}
Choosing $(k_*,s_*)$ for multivariate problems then involves maximizing Eq. \ref{eq:CART}, aggregated across different parameters.
Because this condition, denoted as the CART criterion, only considers the difference in mean parameters in the child nodes and not the whole distributions, \'Cevid et al. developed a second criterion based on the maximal mean discrepancy (MMD) metric \citep{cevid_distributional_2022}.
The MMD defines the difference between distributions of particles in two different sets by employing a positive-definite kernel and its embedding into a Reproducing Kernel Hilbert Space (RKHS) \citep{10.7551/mitpress/7503.003.0069}.
Due to the computational expense of MMD and the need to compute it for many different statistic and threshold candidates, the authors replace the kernel with its Fourier approximation.
By default, they use the Gaussian kernel
$k(\theta,\theta')=\left.\exp\left(-\lVert\theta-\theta'\rVert_2^2\middle/2\sigma^2\right)\middle/\left(\sigma\sqrt{2\pi}\right)^{|\Theta|}\right.$ with parameter count $|\Theta|$ and bandwidth $\sigma$, the median pairwise $L^2$ distance between the parameters of the particles in the parent node.
With Fourier features $\omega_1,\dots,\omega_L$ randomly chosen from the multivariate Gaussian distribution $\text{Normal}_{|\Theta|}\left(\textbf{0},\sigma^{-2}\textbf{I}_{|\Theta|}\right)$ where $\textbf{0}$ and $\textbf{I}_{|\Theta|}$ are square zero and identity matrices of size $|\Theta|\times|\Theta|$, \'Cevid et al. derive the approximate MMD as
\begin{equation}\label{eq:MMD}
    \mathcal{D}(R_{*1},R_{*2})=
    \frac{1}{L}\sum_{l=1}^L\frac{|R_{*1}|\cdot|R_{*2}|}{|R_{*}|^2}\left|\frac{1}{|R_{*1}|}\sum_{\theta\in R_{*1}}\varphi_{\omega_l}(\theta)
    -
    \frac{1}{|R_{*2}|}\sum_{\theta\in R_{*2}}\varphi_{\omega_l}(\theta)
    \right|^2
\end{equation}
where $\varphi_\omega(u)=e^{i\omega^Tu}$ are the approximate kernels.
Constructing the random forest using the MMD criterion (Algorithm \ref{alg:abc-drf-1}) consists of splitting each node, such that (\ref{eq:MMD}) is maximized.

\bigskip
\begin{minipage}{0.8\textwidth}
\begin{algorithm}[H]
    \caption{Growing regression trees for DRF \citep{cevid_distributional_2022}}\label{alg:abc-drf-1}
    Form reference table $R=\left\{\left(\theta^{(1)},s^{(1)}\right),\dots,\left(\theta^{(N)},s^{(N)}\right)\right\}$, similar to Algorithm \ref{alg:abc-rf-1}
    
    \For{$t=1,\dots,B$}{
        $R_{t}\gets$ subsample of $R$ of size $n_{\rm sample}$;
        $R_t$ is split into $R_{t1}$ and $R_{t2}$
    
        Tree $T_t$ is rooted in node $R_{t1}$
        
        \While{$\max_{\mathrm{leaf }R_*\in T_t}|R_*|>N_{\rm min}$}{
            Find a leaf $R_*$ in $T_t$ with $\left|R_*\right|>N_{\rm min}$

            Statistic candidates $\gets$ subsample of $S$ with size $\min(\max(\tilde n_{\rm try},1),|S|)$, where $\tilde n_{\rm try}\sim\text{Poisson}(n_{\rm try})$

            \If{$\mathrm{criterion}=\mathrm{CART}$}{
                Choose statistic $k_*$ and threshold $s_*$ to split node $R_*$ into new leaves $R_{*1}=\left\{(\theta,s)\in R_*:s_{k_*}\le s_*\right\}$ and $R_{*2}=\left\{(\theta,s)\in R_*:s_{k_*}> s_*\right\}$, such that the CART criterion (Eq. \ref{eq:CART}), aggregated across all parameters, is maximized
            }\ElseIf{$\mathrm{criterion}=\mathrm{MMD}$}{
                $\sigma\gets$ median pairwise distance between $\theta$'s in $R_*$

                Fourier features $\omega_1,\dots,\omega_L\sim\text{Normal}_{|S|}\left(\textbf{0},\sigma^{-2}\textbf{I}_{|S|}\right)$

                Choose $k_*$ and $s_*$ to split $R_*$ into $R_{*1}$ and $R_{*2}$, such that the MMD criterion (Eq. \eqref{eq:MMD}), defined with $\omega_l$'s, is maximized
            }
        }

        \For{$i=1,\dots,\left|R_{t2}\right|$}{
            Follow $T_t$ from root node with $s^{(i)}$ to locate leaf $L_t\left(s^{(i)}\right)$
        }
    }
\end{algorithm}
\end{minipage}
\bigskip

Once the random forest is constructed, computing the posterior distribution for an observation $y^{\rm obs}$ consists of finding the leaf $L_t\left(s^{\rm obs}\right)$ in each tree $T_t$ that the observed statistics $s^{\rm obs}$ falls into.
Because the samples for tree construction are not bootstrappped, the weight for each particle $\theta^{(i)}$ in the reference table is simply the normalized number of times that it ends up in the same leaf as $s^{\rm obs}$ across the entire forest (Algorithm \ref{alg:abc-drf-2}).

The CDF of the joint posterior distribution $\pi_{ABC-DRF}\left(\theta\middle|s^{\rm obs}\right)$ is then approximated as
\begin{equation*}
    \mathbb{P}\left(\theta_1\leq x_1,\dots,\theta_{|\Theta|}\leq x_{|\Theta|}\right)=
    \sum_{i=1}^N w^{(i)}\textbf{1}_{\theta^{(i)}_1\leq x_1,\dots,\theta^{(i)}_{|\Theta|}\leq x_{|\Theta|}}
\end{equation*}

\bigskip
\begin{minipage}{0.8\textwidth}
\begin{algorithm}[H]
    \caption{Multivariate distribution from DRF \citep{cevid_distributional_2022}}\label{alg:abc-drf-2}
    \For{$t=1,\dots,B$}{
        Follow the tree $T_t$ from root node with $s^{\rm obs}$ until locating leaf $L_t(s^{\rm obs})$
    }
    \For{$i=1,\dots,N$}{
        Weight for particle $\theta^{(i)}\gets w^{(i)}=\frac{1}{B}\sum_{t=1}^{B}\frac{\textbf{1}_{L_t\left(s^i\right)=L_t\left(s^{\rm obs}\right)}}{\left|L_t\left(s^{\rm obs}\right)\right|}$
    }
\end{algorithm}
\end{minipage}
\bigskip

\cite{cevid_distributional_2022} developed an {\sf R} package \texttt{drf} based on Algorithms \ref{alg:abc-drf-1} and \ref{alg:abc-drf-2}.
Compared to \texttt{abcrf}, the authors employ different methods to maximize forest diversity (Algorithm \ref{alg:abc-drf-1}).
Instead of a fixed number of $n_{\rm try}$ randomly selected statistics from which the splitting condition is chosen, the number of candidate statistics in DRF varies between nodes, following $\min(\max(\tilde n_{\rm try},1),|S|)$ where $\tilde n_{\rm try}$ is sampled from $\text{Poisson}(n_{\rm try})$.
Furthermore, instead of bootstrapping the reference table, the samples $R_{t1}$ to form each tree's root node are subsampled from $R$.
After the tree is constructed, the particles actually stored in the leaves for computing posterior weights are from a disjoint subsample $R_{t2}$ of $R$, in order to minimize overfitting. By default, the total number of particles used for tree construction $R_{t1}$ and weights $R_{t2}$ is $N/2$.

In the next section we describe how DRF may be adapted for use in approximate Bayesian computation.

\section{ABC-DRF}\label{sec:abcdrf}

The main limitation of \texttt{abcrf} is that it is typically used for inference of one parameter at a time, resulting in difficulty studying the \emph{joint} posterior of the elements of $\theta$. We noted in \citep{dtz24} that DRF may be exploited for use in ABC by constructing the reference table just as for ABC-RF, and then using \texttt{drf} to generate observations from the joint posterior.  We give an illustration of this approach, called ABC-DRF, in this section.

\subsection{A  hierachical Normal mean example}\label{sec-hierarchical}

We illustrate ABC-DRF with an adaptation of an example from  \citep{raynal2019abc} and discussed in \citep{dtz24}. The model has 
\begin{eqnarray}
    y_1, y_2, \ldots, y_n \mid \theta_1, \theta_2 & \sim & {\rm Normal}(\theta_1, \theta_2) \label{eq:hierarchical-1}\\
    \theta_1 \mid \theta_2 &\sim &\text{Normal}(0,\theta_2) \nonumber \\
    \theta_2 & \sim & \text{IG}(\alpha,\beta) \nonumber   
\end{eqnarray}
where $\text{IG}(\alpha,\beta)$ denotes the inverse gamma distribution  with shape $\alpha$ and rate $\beta$ and density 
$$
f(x;\alpha,\beta)=\frac{\beta^\alpha}{\Gamma(\alpha)}(1/x)^{\alpha+1}\exp(-\beta/x), \quad x>0.
$$
The observed data $y^{\rm obs}$ consists of $n$ data points $y^{\rm obs}_1,\dots,y^{\rm obs}_n$.

The joint posterior distribution of $(\theta_1,\theta_2)$ is determined by
\begin{eqnarray}
    \theta_2 \mid y^{\rm obs} & \sim &
    \text{IG}\left(\frac{n}{2}+\alpha,\,B\right) \label{eq:hierarchical-truth-1}\\
     \theta_1 \mid \theta_2, y^{\rm obs} & \sim & 
    \text{Normal}\left(\frac{n\cdot\bar{y}^{\rm obs}}{n+1}, \frac{2\cdot\theta_2}{n+1}\right),\label{eq:hierarchical-truth-1a} 
\end{eqnarray}
where
$$
    B=\frac{1}{2}\left((S^{obs})^2 + 2 \beta + \frac{n\cdot(\bar{y}^{\rm obs})^2}{n+1}\right), \ 
\bar{y}^{obs}=n^{-1}\sum_{i=1}^ny_i^{\rm obs},  \ S^{obs}=\sqrt{\sum_{i=1}^n\left(y_i^{\rm obs}-\bar{y}^{\rm obs}\right)^2}.
$$
The marginal posterior of $\theta_1$ is 
\begin{equation*}
    \theta_1 \mid y^{\rm obs}\sim
    \frac{n\cdot\bar{y}^{\rm obs}}{n+1}+
    \sqrt{\frac{2B}{(n+1)(n+2 \alpha)}}\cdot\text{T}_{n+2\alpha}
\end{equation*}
where $\text{T}_m$ denotes the $t$-distribution with $m$ degrees of freedom. $\theta_1$ and $\theta_2$ are uncorrelated under the posterior.

To illustrate the behavior of ABC-DRF, we generated a test set $y^{\rm obs} = (y_1^{\rm obs},\ldots,y_{10}^{\rm obs})$ from the model, and let $s^{\rm obs}=S(y^{\rm obs})$ consist of the 61 summary statistics described in \cite{raynal2019abc}.
The first three statistics $s^{\rm obs}_1,s^{\rm obs}_2,s^{\rm obs}_3$ are the mean, variance and median absolute deviation of $y^{\rm obs}$.
The next eight statistics $s^{\rm obs}_4,\dots,s^{\rm obs}_{11}$ are sums and products of either two or all values of $s^{\rm obs}_1,s^{\rm obs}_2,s^{\rm obs}_3$.
The final 50 statistics are noise: $s^{\rm obs}_{12},\dots,s^{\rm obs}_{61}\sim$ Uniform$(0,1)$.

We first infer the posterior distribution from 
\texttt{drf}, following the example in \cite{dtz24}.
The reference table consists of $N=20,000$ entries, each of which results from sampling $\theta$ and  $y_1,\dots,y_n$ from  (\ref{eq:hierarchical-1}), then computing $s$ similarly to $s^{\rm obs}$.
The algorithm \texttt{drf} then infers the joint posterior distribution for $\theta$, which we compare against the true posterior distribution. 

\begin{figure}[H]
    \centering
    \includegraphics[width=16cm]{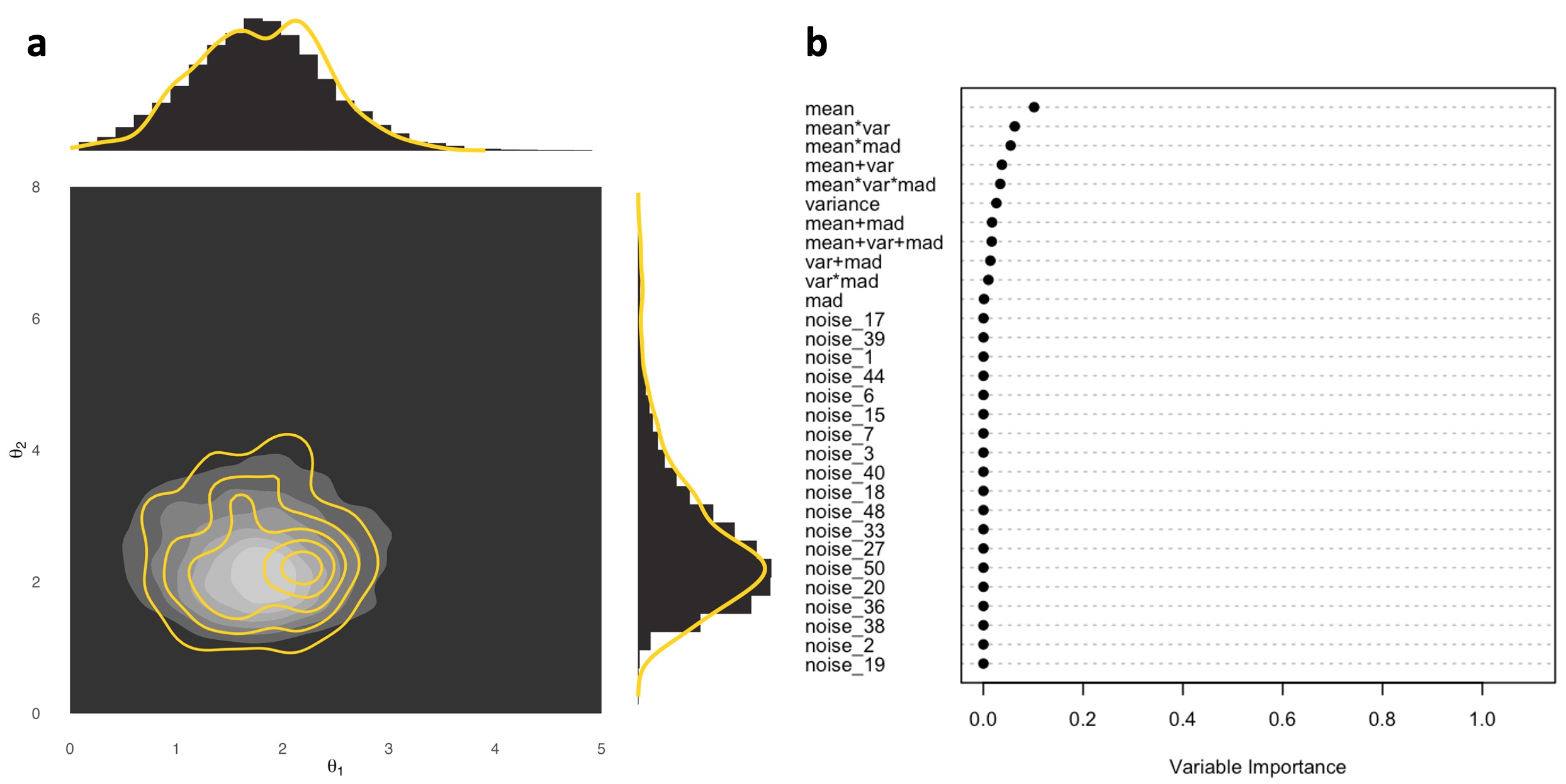}
    \caption{
    Inference of $\theta=(\theta_1,\theta_2)$ in the hierarchical model, with $\alpha=4,\beta=5$.
    \textbf{a}: 
    Joint posterior distributions for $\theta_1$ and $\theta_2$,
    inferred from ABC-DRF with $N=20,000$ simulations (yellow contours)
    against ground truth (density heatmap in gray-scale), with marginal distributions for each parameter from ABC-DRF (yellow line) against ground truth (black histogram)
    \textbf{b}: 
    Variable importance analysis for summary statistics from ABC-DRF, performed with $N=1,000$ simulations.
    }
    \label{fig:hierarchical}
\end{figure}

Figure \ref{fig:hierarchical}a shows the comparison for $\alpha=4,\beta=5$.
The posterior distributions from ABC-DRF are in agreement with the true density.
Furthermore, the variable importance analysis detects information in $s_1,\dots,s_{11}$, which indeed contain signals for the distribution of $y^{\rm obs}$ (Figure \ref{fig:hierarchical}b).
In contrast, the pure noise statistics $s_{12},\dots,s_{61}$ are deemed unimportant, as expected.

Compared to \texttt{abcrf}, \texttt{drf}'s ability to infer the joint distribution is crucial for inference problems where the parameters are known to be dependent. 
We will experiment with such models in Section \ref{sec:results}.

\section{ABC Sequential Monte Carlo with random forests}\label{sec:abcsmcrf}

Despite their promise as relatively simple ABC approaches, there are some drawbacks in using \texttt{abcrf} and \texttt{drf} in practice.
First, their performance typically improves as the forest size increases.
The number of splitting criterion computations in growing a forest of $B$ balanced trees from a reference table of size $N$, where each split considers $n_{\rm try}$ statistics, is $\mathcal{O}(B\cdot n_{\rm try}\cdot N\cdot \log N)$.
For complex problems involving many statistics and requiring large reference tables, the forest can become too computationally expensive to construct.
Second, similar to other ABC implementations, the accuracy of the estimated posterior distributions depends on the number of simulations $N$.
However, because the number of splits is approximately $N \log N$ in balanced trees and more in unbalanced trees, the application of \texttt{abcrf} or \texttt{drf} in difficult inference problems where many simulations are necessary can demand more memory than available resources.

To alleviate these problems, we propose integrating the random forest methods within the ABC sequential Monte Carlo (ABC-SMC) framework (Algorithm \ref{alg:ABC-SMC}), which was introduced by \cite{toni_approximate_2008} as a novel approach to solve the problem of low acceptance rates in the ABC rejection (ABC-REJ) method.
ABC-SMC refines the particle population drawn from the prior distribution iteratively, where the posterior distribution from iteration $t-1$ becomes the prior distribution for iteration $t$.
A common choice for the sampling distribution in iteration $t$ is $g_t(\theta)=\sum_{k=1}^{N_{t-1}}w_{t-1}^{*(k)}\cdot K_t\left(\theta_t^{(k)}\middle|\theta_{t-1}^{(k)}\right)$, which samples a random particle from iteration $t-1$ then perturbs it with a Markov kernel $K_t(\theta | \theta')$ \citep{sisson2018handbook}.
The statistics simulated from this new particle are compared against data statistics with a kernel function $H_h$ centered at 0.
The kernel scale parameters decrease in successive iterations (i.e., $h_1\ge\dots\ge h_T$), hence only particles increasingly closer to $s_{obs}$ are retained.
The thresholds $c_2,\dots,c_T$ further control the acceptance or rejection and resampling of particles in each iteration.
The final set of particles $\left\{\theta_T^{(1)},\dots,\theta_T^{(N)}\right\}$ from iteration $T$ is drawn from $\pi_{\text{ABC-SMC}}\left(\theta \middle| s^{\text{obs}} \right)\propto\displaystyle \int H_{h_T}\left( \left\| s_t^{(i)} - s^{\text{obs}} \right\| \right)\cdot p\left( s \middle| \theta \right)\cdot \pi(\theta) \, ds$, where $p\left(s\middle|\theta\right)$ is the probability distribution of observed statistics $s$ from the model, given parameter $\theta$ \citep{sisson2018handbook}.

\begin{minipage}{0.8\textwidth}
\begin{algorithm}[H]
    \caption{ABC-SMC (adapted from Alg. 4.7, page 111, \citep{sisson2018handbook})}\label{alg:ABC-SMC}
    \SetKwInOut{Input}{Input}
    \SetKwInOut{Output}{Output}
    \For{$t=1, \ldots, T$}{
        \If{$t=1$}{
            \For{$i=1,\dots,N_t$}{
                Generate $\theta_1^{(i)}\sim g_1(\theta)$, an initial sampling distribution

                Simulate data $y_1^{(i)}$ with $\theta_1^{(i)}$, and compute $s_1^{(i)}=S\left(y_1^{(i)}\right)$

                Compute weight $w_1^{*(i)}=\left.H_{h_1}\left(\left\|s_1^{(i)}-s^{\rm obs}\right\|\right)\pi\left(\theta_1^{(i)}\right)\middle/g_1\left(\theta_1^{(i)}\right)\right.$
            }
        }
        \Else{
            Construct sampling distribution $g_t(\theta)$ from $\left\{\left(\theta^{(k)}_{t-1},w^{*(k)}_{t-1}\right)\right\}$

            \For{$i=1,\dots,N_t$}{
                Generate $\theta_t^{(i)}\sim g_t(\theta)$

                Simulate data $y_t^{(i)}$ with $\theta_t^{(i)}$, and compute $s_t^{(i)}=S\left(y_t^{(i)}\right)$

                Compute weight $w_t^{(i)}=\left.H_{h_t}\left(\left\|s_t^{(i)}-s^{\rm obs}\right\|\right)\pi\left(\theta_t^{(i)}\right)\middle/g_t\left(\theta_t^{(i)}\right)\right.$

                Reject $\theta_t^{(i)}$ with probability $1-r_t^{(i)}$ with $r_t^{(i)}=\min\left(1,\frac{w_t^{(i)}}{c_t}\right)$, and go to Step 10

                Otherwise, accept $\theta_t^{(i)}$ and set modified weight $w_t^{*(i)}=w_t^{(i)}/r_t^{(i)}$
            }
        }
    }
\end{algorithm}
\end{minipage}
\bigskip

We develop ABC sequential Monte Carlo with random forests (ABC-SMC-(D)RF), which incorporate \texttt{abcrf} and \texttt{drf} into ABC-SMC.
The first version, ABC-SMC-DRF (Algorithm \ref{alg:ABC SMC-RF multi}), is applicable for problems with multiple parameters.
The first iteration samples the parameter sets directly from the prior distribution, then applies ABC-DRF on the reference table $R\in\mathbb{R}^{N_1\times(|S|+|\Theta|)}$ that combines all parameter sets and corresponding statistics.
In subsequent iterations, parameter sets are drawn from particles in the previous iteration, with weights estimated from ABC-DRF for $s^{\rm obs}$.
The parameter sets are then perturbed with a multivariate Markov kernel, before ABC-DRF is similarly applied.
Finally, the ABC-DRF weights are re-calibrated (Step 15, Algorithm \ref{alg:ABC SMC-RF multi}), similar to ABC-SMC (Step 12, Algorithm \ref{alg:ABC-SMC}).

\bigskip
\begin{minipage}{0.8\textwidth}
\begin{algorithm}[H]
    \caption{ABC-SMC-DRF for multiple parameters}\label{alg:ABC SMC-RF multi}
    \SetKwInOut{Input}{Input}
    \SetKwInOut{Output}{Output}
    \For{$t=1, \ldots, T$}{
            \For{$i=1,\dots,N_t$}{
                \If{$t=1$}{
                    Generate $\theta_1^{(i)} \sim \pi(\theta)$
                }
                \Else{
                    Sample $\theta^*$ from $\left\{\left(\theta^{(k)}_{t-1},w^{*(k)}_{t-1}\right)\right\}_{k=1,\dots,N_{t-1}}$

                    Generate $\theta_t^{(i)}\sim K_t(\theta | \theta^{*})$

                    If $\pi\left(\theta_t^{(i)}\right)=0$, then return to Step 6
                }
    
                Simulate data $y_t^{(i)}$ with $\theta_t^{(i)}$, and compute $s_t^{(i)}=S\left(y_t^{(i)}\right)$
            }
            
        Form $R=\left\{\left(\theta_t^{(1)},s_t^{(1)}\right),\dots,\left(\theta_t^{(N_t)},s_t^{(N_t)}\right)\right\}$
    
        Perform DRF (Algorithms \ref{alg:abc-drf-1} and \ref{alg:abc-drf-2}) with reference table $R$ to compute weights $w_t^{(1)},\dots,w_t^{(N_t)}$ for observed statistics $s^{\rm obs}$

        \If{$t=1$}{
            $w_t^{*(i)}=w_t^{(i)}$
        }
        \Else{
            $w_t^{*(i)}=w_t^{(i)}\frac{\pi\left(\theta_t^{(i)}\right)}{\sum_{k=1}^{N_{t-1}}w_{t-1}^{*(k)}\cdot K_t\left(\theta_t^{(i)}\middle|\theta_{t-1}^{(k)}\right)}$
        }

        Normalize $\left\{w_t^{*(i)}\right\}$
    }
\end{algorithm}
\end{minipage}
\bigskip

The second version, ABC-SMC-RF (Algorithm \ref{alg:ABC SMC-RF single}), is built around ABC-RF, and designed for univariate problems or finding marginal distributions for multivariate models.
Each parameter in an iteration is sampled independently from the previous iteration and then perturbed.
The statistics are then computed for the whole parameter set.
The reference table consists of these statistics and values corresponding to a given parameter, from which ABC-RF predicts the parameter's marginal distribution based on $s^{\rm obs}$.
The ABC-RF weights are then re-calibrated and normalized.

\bigskip
\begin{minipage}{0.8\textwidth}
\begin{algorithm}[H]
    \caption{ABC-SMC-RF for single parameters}\label{alg:ABC SMC-RF single}
    \SetKwInOut{Input}{Input}
    \SetKwInOut{Output}{Output}
    \For{$t=1, \ldots, T$}{
            \For{$i=1,\dots,N_t$}{
                \If{$t=1$}{
                    Generate $\theta_1^{(i)} \sim \pi(\theta)$
                }
                \Else{
                    \For{$j=1,\dots,|\Theta|$}{
                        Sample $\theta^*_j$ from $\left\{\left(\theta^{(k)}_{j,t-1},w^{*(k)}_{j,t-1}\right)\right\}_{k=1,\dots,N_{t-1}}$

                        Generate $\theta_{j,t}^{(i)}\sim K_{j,t}(\theta | \theta_j^{*})$

                        If $\pi_j\left(\theta_{j,t}^{(i)}\right)=0$, then return to Step 7
                    }

                    $\theta_{t}^{(i)}\gets\left(\theta_{1,t}^{(i)},\dots,\theta_{|\Theta|,t}^{(i)}\right)$
                }
    
                Simulate data $y_t^{(i)}$ with $\theta_t^{(i)}$, and compute $s_t^{(i)}=S\left(y_t^{(i)}\right)$
            }

        \For{$j=1,\dots,|\Theta|$}{
            Form $R_j=\left\{\left(\theta_{j,t}^{(1)},s_t^{(1)}\right),\dots,\left(\theta_{j,t}^{(N_t)},s_t^{(N_t)}\right)\right\}$

            Perform ABC-RF (Algorithms \ref{alg:abc-rf-1} and \ref{alg:abc-rf-2}) with reference table $R_j$ to compute weights $w_{j,t}^{(1)},\dots,w_{j,t}^{(N_t)}$ for observed statistics $s^{\rm obs}$

            \If{$t=1$}{
                $w_{j,t}^{*(i)}=w_{j,t}^{(i)}$
            }
            \Else{
                $w_{j,t}^{*(i)}=w_{j,t}^{(i)}\frac{\pi_j\left(\theta_{j,t}^{(i)}\right)}{\sum_{k=1}^{N_{t-1}}w_{j,t-1}^{*(k)}\cdot K_{j,t}\left(\theta_{j,t}^{(i)}\middle|\theta_{j,t-1}^{(k)}\right)}$
            }

            Normalize $\left\{w_{j,t}^{*(i)}\right\}$
        }
    }
\end{algorithm}
\end{minipage}
\bigskip

The kernel values $H_h\left(\left\|s-s^{\rm obs}\right\|\right)$ in ABC-SMC (Algorithm \ref{alg:ABC-SMC}) increase as $s\rightarrow s^{\rm obs}$.
In the ABC-SMC-(D)RF algorithms, these kernels are replaced by the weights from ABC-DRF in Algorithm \ref{alg:ABC SMC-RF multi} and ABC-RF in Algorithm \ref{alg:ABC SMC-RF single}.
These weights behave similarly to the ABC-SMC kernels: particles with statistics $s$ closer to $s^{\rm obs}$ fall into the same leaves as the data more often across the forest, therefore the corresponding parameters $\theta$ are more heavily weighted.
Furthermore, in ABC-SMC, smaller scale parameters $h_t$ reduce the kernel diffusion, that is, only particles with $\left\|s-s^{\rm obs}\right\|$ increasingly close to 0 have high weights.
Decreasing the minimum node sizes $N_{min}$ in ABC-(D)RF has the same effect, as it reduces the accepted particles across the forest to only those with statistics very close to those of the data.
Finally, ABC-SMC-(D)RF does not reject particles before (D)RF is performed in each iteration except for those outside the prior distribution, and hence is analogous to ABC-SMC with rejection thresholds $c_2=\dots=c_T=0$.

We have developed an {\sf R} package \texttt{abcsmcrf} based on Algorithms \ref{alg:ABC SMC-RF multi} and \ref{alg:ABC SMC-RF single}.
It offers several practical advantages when compared to direct applications of \texttt{abcrf} and \texttt{drf}. First, constructing iterative random forests of size $B$ from reference tables of sizes $N_1,\dots,N_T$ requires $\mathcal{O}\left(B\cdot\sum_{t=1}^T(N_t\cdot\log N_t)\right)$ splits, assuming balanced trees.
This requires less computational resources than growing a random forest based on a reference table of size $N=\sum_{i=1}^TN_i$, which contains $\mathcal{O}\left(B\cdot
N \cdot \log N)\right)$ splits.
Therefore, with the same total number of model simulations, ABC-SMC-(D)RF requires less memory and computational runtime.
Second, similar to the behavior of ABC-SMC compared to ABC-REJ, ABC-SMC-(D)RF is likely to converge to the posterior distribution faster than ABC-RF or ABC-DRF, because the parameter distributions are constantly updated to focus on regions in the parameter space $\Theta$ that best explain $s^{\rm obs}$.
We will test this with some examples in the next section.
Finally, unlike ABC-RF and ABC-DRF which require the full reference table before forest construction and parameter prediction, the approximated posterior distribution from each iteration of ABC-SMC-(D)RF can be compared against the previous iteration to assess whether it has converged, potentially further lowering computational expense.

\section{Results}\label{sec:results}
In this section, we compare the performance of ABC-SMC-(D)RF (Algorithm \ref{alg:ABC SMC-RF multi} or \ref{alg:ABC SMC-RF single}) across different inference problems, against 
ABC-SMC (Algorithm \ref{alg:ABC-SMC}),  ABC-RF (Algorithms \ref{alg:abc-rf-1}, \ref{alg:abc-rf-2}) and ABC-DRF (Algorithms \ref{alg:abc-drf-1}, \ref{alg:abc-drf-2}).
Unless specified otherwise, for ABC-RF, ABC-DRF and  ABC-SMC-(D)RF, we use the default parameters in \texttt{abcrf} and \texttt{drf} for tree count $B$, candidate statistic count $n_{\rm try}$, leaf size threshold $N_{\rm min}$, root node subsample size $n_{\rm sample}$, etc.

\subsection{Deterministic Lotka-Volterra model}

The Lotka-Volterra model describes the interaction dynamics between predators and prey 
\citep{Lotka_elements_1925,volterra_variations_1928}.
The deterministic model for the number of prey $x$ and predators $y$ is in the form of paired nonlinear differential equations:
\begin{equation*}
    \begin{aligned}
    \frac{\mathrm{d} x}{\mathrm{d} t}&=ax-cxy \\
    \frac{\mathrm{d} y}{\mathrm{d} t}&=bxy-dy
    \end{aligned}    
\end{equation*}
where $a$ is the prey's birth rate, $d$ is the predator's death rate, and $c$ and $b$ are the prey's death rate and predator's birth rate due to predation, respectively.
We fix $c=d=1$ and seek to infer $\theta=(a,b)$ from prior distribution $a,b\sim\text{Uniform}(-10,10)$.
Similar to \cite{toni_approximate_2008}, we solve the ODE system with initial condition $(x(0), y(0))=(1, 0.5)$ with $\theta=(1,1)$, and sample $(x,y)$ at eight time points between $t=0$ and $t=15$, then add Normal$(0,0.5^2)$ noise to each data point to form the observed statistics $s^{\rm obs}=\left\{x_1,y_1,\dots,x_8,y_8\right\}$.

\begin{figure}[htb]
    \centering
    \includegraphics[width=16cm]{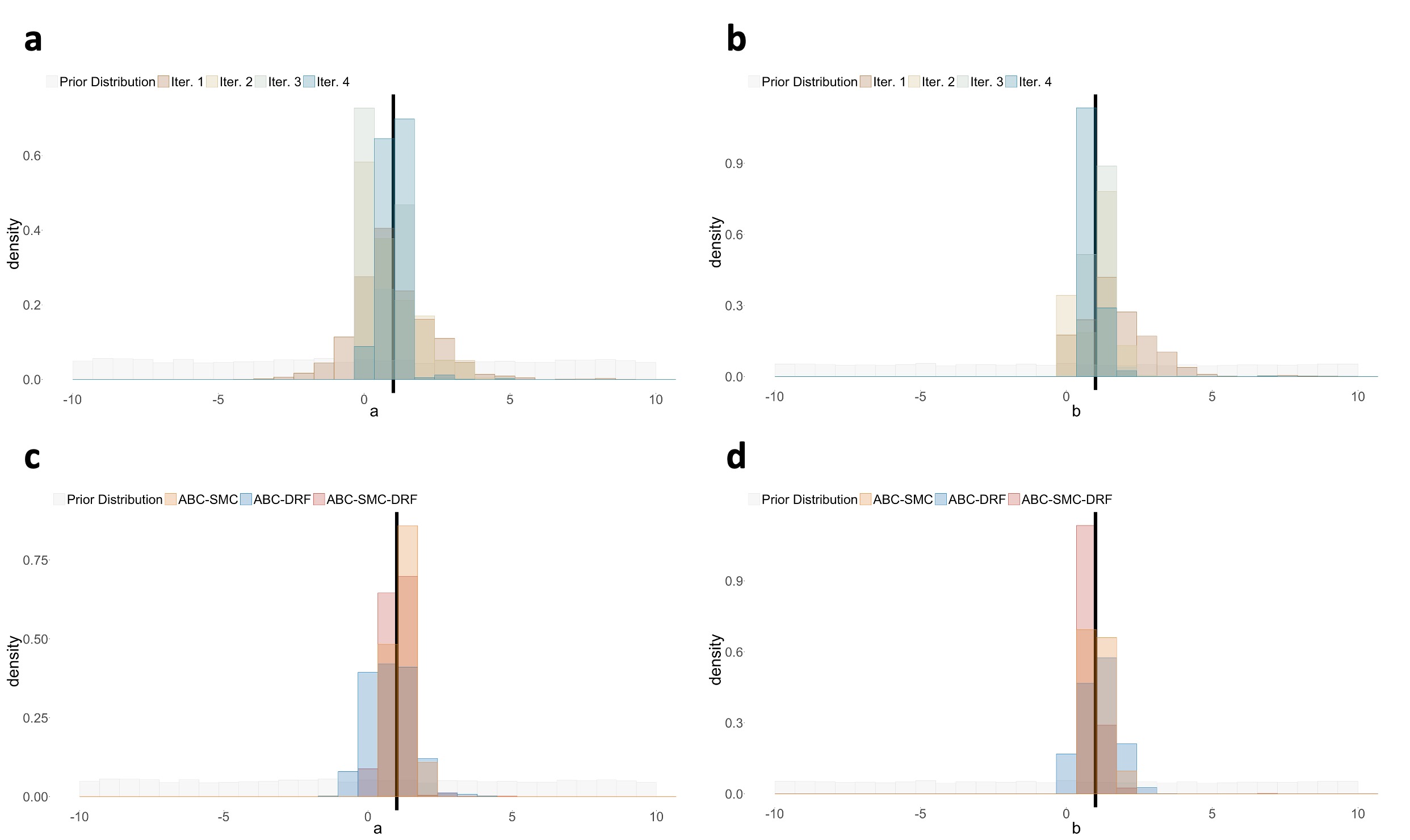}
    \caption{
    Parameter inference for the deterministic Lotka-Volterra model.
    \textbf{a, b}: Iterative posterior distributions for reaction rates $a$ (\textbf{a}) and $b$ (\textbf{b}) from ABC-SMC-RF.
    \textbf{c, d}: Marginal posterior distributions for $a$ (\textbf{c}) and $b$ (\textbf{d}), from ABC-SMC, ABC-DRF and ABC-SMC-DRF.
    Black vertical lines denote true parameter values.
    }
    \label{fig:lv}
\end{figure}

We first implement ABC-SMC-(D)RF for multiple parameters.
Algorithm \ref{alg:ABC SMC-RF multi} is applied with $T=4$ iterations, each with $N_t=5,000$ simulations in the reference table, and perturbation kernels $K_t\left(\theta\middle|\theta^*\right)=\theta^*+\text{Uniform}(-0.1,0.1)$.
Figures \ref{fig:lv}a, b show the distributions of $a$ and $b$ after each iteration $t$ in \texttt{abcsmcrf}.
The distributions become more concentrated as $t$ progresses, but stay centered around the true values of $(a,b)$ (Table \ref{table:lv}).

To evaluate \texttt{abcsmcrf}'s performance, we compare the final posterior distributions with ABC-DRF and ABC-SMC.
Algorithm \texttt{drf} is performed on a reference table of size $N=20,000$, to match the total number of simulations in \texttt{abcsmcrf}.
We use the {\sf R} package \texttt{EasyABC} \citep{easyabc-package} to implement ABC-SMC with $T=5$ iterations.
\texttt{EasyABC} uses the distance $d\left(s,s^{\rm obs}\right)=\sum_{i=1}^8\left[(x_i-x_i^{\rm obs})^2+(y_i-y_i^{\rm obs})^2\right]$ to compare simulated statistics $s$ against observation $s^{\rm obs}$, and adaptive perturbation kernels $K_t\left(\theta\middle|\theta'\right)=\text{Normal}\left(\theta',2\cdot\text{var}\left(\theta_{t-1}\right)\right)$ as proposed in \cite{beaumont_adaptive_2009}.
Referencing \cite{toni_approximate_2008}, we impose the series of tolerance thresholds $\epsilon_1 =30.0, \epsilon_2=16.0, \epsilon_3=6.0, \epsilon_4=5.0, \epsilon_5=4.3$.
To produce 1,000 accepted particles, \texttt{EasyABC} requires $N=56,850$ simulations.

Figures \ref{fig:lv}c, d present the posterior distributions approximated by the ABC implementations.
The marginal distributions from ABC-SMC-DRF
are as centered and concentrated around the true values for $(a,b)$ as ABC-SMC (Table \ref{table:lv}).
In contrast, even though the distributions from DRF are centered around the ground truth, the variances are consistently higher, indicating higher uncertainty.

\begin{table}[!h]
    \begin{center}
    \begin{tabular}{|c|c|c|c|c|c|c|}
        \hline
        \multirow{2}{*}{Statistics} & \multicolumn{4}{c|}{ABC-SMC-DRF iterations} & \multirow{2}{*}{ABC-DRF} & \multirow{2}{*}{ABC-SMC} \\
        \cline{2-5}
        & 1 & 2 & 3 & 4 (Final posterior) &  &  \\[0.5ex] 
        \hline\hline
        $\mathbb{E}(a)$ & 0.8237 & 0.3349 & 1.2044 & \textbf{1.1215} & 0.7562 & 1.2912\\
        \hline
        Var(a) & 1.9018 & 0.2757 & 0.1279 & \textbf{0.0333} & 0.5953 & 0.1104\\
        \hline
        $\mathbb{E}(b)$ & 1.6816 & 1.4889 & 1.0061 & 0.9704 & 1.3092 & \textbf{1.0269}\\
        \hline
        Var(b) & 1.5298 & 0.1536 & 0.0697 & \textbf{0.0313} & 0.3593 & 0.1049\\
        \hline
    \end{tabular}
    \caption{
    Means and variances of marginal posterior distributions for the deterministic Lotka-Volterra model from ABC-SMC-DRF, ABC-DRF, and ABC-SMC.
    The best result for each statistic across different algorithms is in bold ($\mathbb{E}(a)$, $\mathbb{E}(b)$ closest to true values $(a,b)=(1,1)$, and lowest variance in each statistic).
    }
    \label{table:lv}
    \end{center}
\end{table}

\subsection{Stochastic biochemical reaction systems}

We examine ABC-SMC-DRF's application in inferring reaction rates for biochemical systems.
We first examine the Michaelis-Menten model, which describes the kinetics between an enzyme $E$ that binds to a substrate $S$ to form a complex $ES$, leading to product $P$:
\begin{align}
    ES&\rightarrow E+P \label{eq:mm-1}\\
    E+S&\rightarrow ES \label{eq:mm-2}\\
    ES&\rightarrow E+S \label{eq:mm-3}
\end{align}
The state of the system at time $t$ can be described as $X(t)=\left[E(t),S(t),ES(t),P(t)\right]$, the counts of each type of molecule.
The time for the next reaction (\ref{eq:mm-1}), (\ref{eq:mm-2}) and (\ref{eq:mm-3}) to occur is exponentially distributed, with rates defined by their propensity functions:
\begin{align*}
    \alpha_1(X(t))&=\bar c_1\cdot ES(t);\quad \bar c_1=c_1\\
    \alpha_2(X(t))&=\bar c_2\cdot E(t)\cdot S(t);\quad \bar c_2=\frac{10^{c_2}}{\text{nA}\cdot\text{vol}}\\
    \alpha_3(X(t))&=\bar c_3\cdot ES(t);\quad \bar c_3=10^{c_3}
\end{align*}
respectively, where the Avagadro's constant nA $=6.023\cdot 10^{23}$ approximates the number of molecules in a mole, vol $=10^{-15}$ is the volume of the system, and $\theta=(c_1,c_2,c_3)$ parameterizes the Michaelis-Menten model for different biochemical systems \citep[Example 7.3]{wilkinson2018stochastic}.

The initial state is defined by $E(0)=2\cdot 10^{-7}\cdot\text{nA}\cdot\text{vol}$, $S(0)=5\cdot 10^{-7}\cdot\text{nA}\cdot\text{vol}$, and $ES(0)=P(0)=0$ \citep{wilkinson2018stochastic}.
We seek to infer $\theta$ from $s^{\rm obs}=\left\{X(t),t=1,\dots,10\right\}$, simulated with true parameters $c_1=0.1,c_2=6,c_3=-4$.
We compare the results from \texttt{drf} with $N=$ 20,000 simulations, and \texttt{abcsmcrf} with $T=$ 5 iterations, each with reference tables of size $N_t=$ 4,000, from uniform prior distributions $\pi\left(c_1\right)=$ U$(0,1)$, $\pi\left(c_2\right)=$ U$(5,7)$, $\pi\left(c_3\right)=$ U$(-5,-3)$.
The perturbation kernels in \texttt{abcsmcrf} in each iteration are $K_t\left(c_1\middle|c_1'\right)=c_1'+\text{U}(-0.05,0.05)$, $K_t\left(c_2\middle|c_2'\right)=c_2'+\text{U}(-0.1,0.1)$, and $K_t\left(c_3\middle|c_3'\right)=c_3'+\text{U}(-0.1,0.1)$.
All simulations are generated by applying Gillespie's algorithm \citep{gillespie1977exact}.

\begin{figure}[H]
    \centering
    \includegraphics[width=16cm]{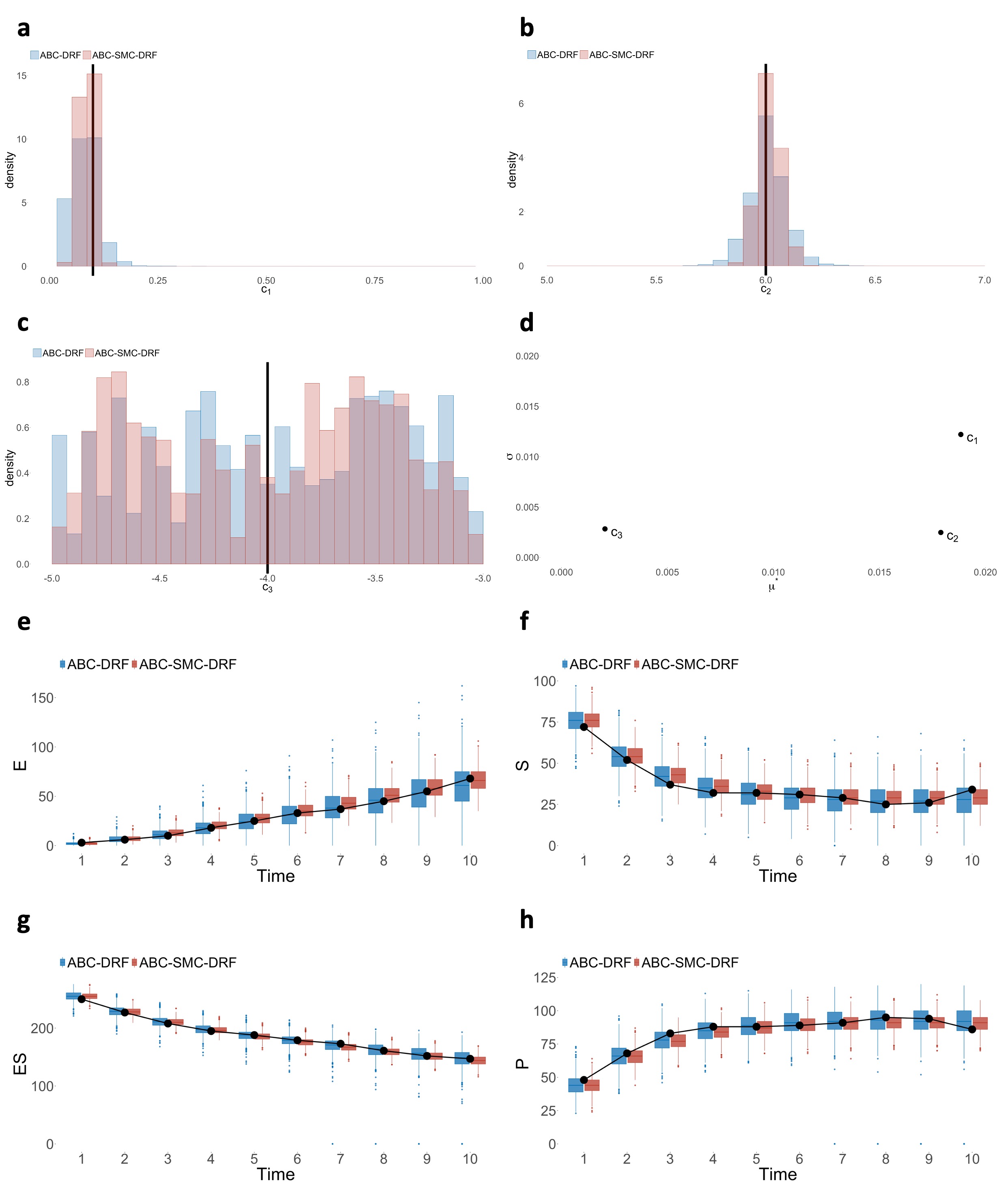}
    \caption{
    Parameter inference for the Michaelis-Menten reaction system.
    \textbf{a, b, c}:
    Marginal distributions for reaction rates $c_1$ (\textbf{a}), $c_2$ (\textbf{b}) and $c_3$ (\textbf{c}), inferred with ABC-DRF and ABC-SMC-DRF, compared against true values (black lines).
    \textbf{d}: 
    Absolute mean ($\mu^*$) and standard deviation ($\sigma$) of elementary effects of $c_1,c_2$ and $c_3$, using Morris's global sensitivity analysis.
    \textbf{e, f, g, h}:
    Range of molecule numbers for $E$ (\textbf{e}), $S$ (\textbf{f}), $ES$ (\textbf{g}) and $P$ (\textbf{h}) across time, simulated with $\theta\sim\pi_{ABC-DRF}\left(\theta\middle|s^{\rm obs}\right)$ and $\pi_{ABC-SMC-DRF}\left(\theta\middle|s^{\rm obs}\right)$ (box plots), compared against observed data $s^{\rm obs}$ (black dots and lines).
    }
    \label{fig:mm}
\end{figure}

Comparing the results for $c_1$ and $c_2$ shows that ABC-SMC-DRF's posterior distributions ($c_1$: mean = 0.110, 95\% CI = $[0.083, 0.141]$; $c_2$: mean = 5.959, 95\% CI = $[5.905, 6.016]$) are centered closely around the true values of $c_1=0.1,c_2=6$, similarly to ABC-DRF, but with narrower confidence intervals (ABC-DRF $c_1$: mean = 0.102, 95\% CI = $[0.023, 0.176]$; $c_2$: mean = 5.943, 95\% CI = $[5.759, 6.093]$) (Figure \ref{fig:mm}a, b).
In contrast, both methods are uncertain about the true distribution for $c_3$ (ABC-SMC-DRF: mean = -3.994, 95\% CI = $[-4.887, -3.259]$; ABC-DRF: mean = -3.950, 95\% CI = $[-4.915, -3.032]$) (Figure \ref{fig:mm}c).

Analyzing the reaction system reveals a likely reason: assuming true values of $\theta$, reactions (\ref{eq:mm-1}) and (\ref{eq:mm-3}) occur with rates $\bar c_1=0.1$ and $\bar c_3=10^{-4}$, respectively. Because of the big difference in magnitudes, complex $ES$ overwhelmingly undergoes reaction (\ref{eq:mm-1}), while reaction (\ref{eq:mm-3}) rarely occurs and does not significantly impact the observed molecule counts.

We apply Morris's global sensitivity analysis \citep{morris1991factorial} to study the parameter identifiability (R package \texttt{sensitivity} \citep{pkg-sensitivity, monari2017characterization}).
The method involves computing the elementary effect of each parameter, defined in this case as the change in the statistics divided by the change in the parameter, across different sampling schemes, and analyzing its absolute mean $\mu^*$ and standard deviation $\sigma$.
The study shows that $c_1$ ($\mu^*=0.019, \sigma=0.012$) and $c_2$ ($\mu^*=0.018, \sigma=0.002$) have significantly higher absolute mean elementary effects than $c_3$ ($\mu^*=0.002, \sigma=0.003$) (Figure \ref{fig:mm}d).
This implies that varying $c_3$ has little to no effect on the observed statistics, consistent with earlier findings \citep{degasperi2007sensitivity}.

Indeed, despite the uncertainty in $c_3$, simulations performed with $\theta$ values sampled from the posterior distributions of ABC-DRF and ABC-SMC-DRF are centered around the observed values $s^{obs}$ across different molecules and time points (Figure \ref{fig:mm}e, f, g, h).
Moreover, consistent with ABC-SMC-DRF's higher certainty in the inference for $c_1$ and $c_2$, molecule counts simulated from its posterior distribution have significantly reduced range compared to ABC-DRF.

\subsection{Stochastic birth-death models}\label{section:bd_initial_parameters}

\subsubsection{Linear birth-death process}\label{lbpsect}

We next examine the performance of ABC-SMC-DRF for the homogeneous linear birth-death process illustrated in \cite{tavare2018linear}, for which exact results may be found. 
The model starts at time 0 with $Z(0) = z_0$ individuals, each of which divides at rate $\lambda>0$ and dies at rate $\mu>0$, independently for each individual. 
The number of individuals at time $t>0$ is denoted $Z(t)$, with expected value $\mathbb{E}[Z(t)] = Z(0) \mathrm{e}^{(\lambda-\mu) t}$. 

The probability $\mathbb{P}(Z(t)=j \mid Z(0)=i)$ that there are $j$ individuals at time $t$, given that there are $i$ individuals at time $0$, follows from \cite{nk1975} as
\begin{equation}\label{Zstdistn}
    \mathbb{P}(Z(t)=j \mid Z(0)=i) =\sum_{l=0}^{\min(i,j)} \binom{i}{l} (1 - \alpha(t))^l \alpha(t)^{i-l} \cdot \binom{j-1}{i-l}(1-\beta(t))^l \beta(t)^{j-l}, j = 1, 2, \ldots
\end{equation}
and
\begin{equation}\label{Zstdistn0}
  \mathbb{P}(Z(t)= 0 \mid Z(0)=i)  = \alpha(t)^i,
\end{equation}
where $\alpha(t)$ and $\beta(t)$ are defined by 
\begin{equation*}
    \begin{aligned}
        \alpha(t) = \frac{\mu \left( e^{(\lambda - \mu) t} - 1 \right)}{\lambda e^{(\lambda - \mu) t} - \mu},\quad\beta(t)= \frac{\lambda}{\mu} \alpha(t),&\quad\text{ if }\lambda\neq\mu;\\
        \alpha(t)=\beta(t)=\frac{\lambda t}{1+\lambda t},&\quad\text{ if }\lambda=\mu,
    \end{aligned}
\end{equation*}
\citep{kendall1948generalized}. The Markov property implies that the likelihood for $\lambda$ and $\mu$, given an observation $y^{\rm obs} = (Z(t_1)=z_1,\dots,Z(t_n)=z_n)$ with $z_n > 0$, is
\begin{equation}\label{eq:bd-likelihood}
    f\left(\lambda,\mu\middle|y^{\rm obs}\right)=\frac{\prod_{i=0}^{n-1}\mathbb{P}(Z(t_{i+1})=z_{i+1}\mid Z(t_i)=z_i)}{\mathbb{P}(Z(t_n) > 0 \mid Z(0) = z_0)},
\end{equation}
which may be calculated from (\ref{Zstdistn}) and (\ref{Zstdistn0}). The observations used in the example below are generated on a grid of time points $t_1 = 1/25, t_2 = 2/25, \ldots, t_{25} = 1,$ and $z_0 = 10.$

\begin{figure}[H]
    \centering
    \includegraphics[width=16cm]{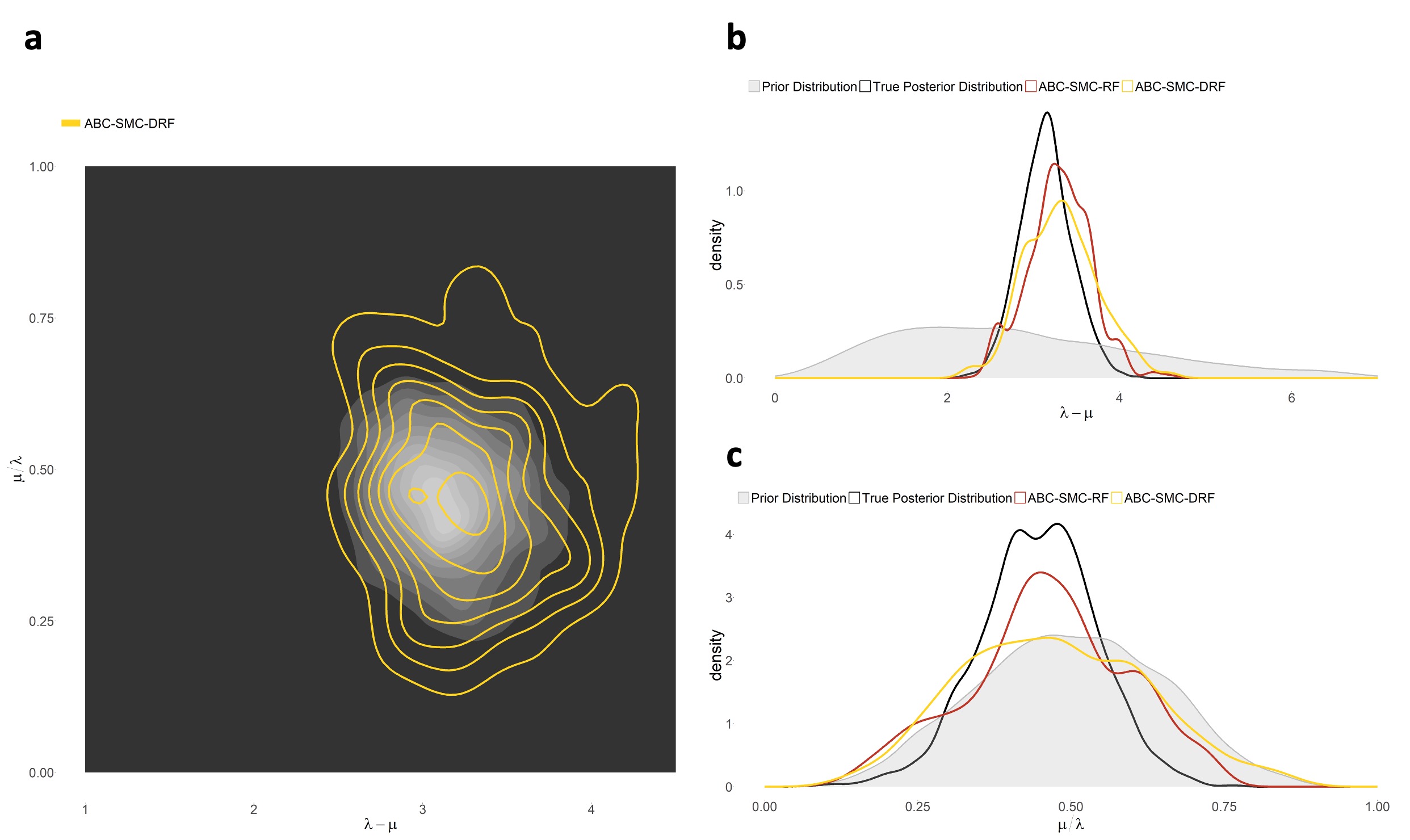}
    \caption{
    Parameter inference for the linear bith-death branching process.
    \textbf{a}:
    Joint posterior distributions for the Malthusian  parameter $\lambda - \mu$ and the extinction parameter $\mu/\lambda$  from ABC-SMC-DRF (yellow contours) against ground truth (Eq. \ref{eq:bd-likelihood}, density heatmap in gray-scale).
    \textbf{b, c}:
    Marginal distributions for $\lambda - \mu$ (\textbf{b}) and $\mu/\lambda$ (\textbf{c}), inferred with ABC-SMC-RF and ABC-SMC-DRF.
    }
    \label{fig:bd_reparametrized}
\end{figure}

Instead of approximating the posterior distribution of $\theta = (\lambda, \mu)$, we approximate that of the extinction parameter $\mu/\lambda$ and the Malthusian parameter $\lambda - \mu$, so that $\theta = (\mu/\lambda, \lambda - \mu)$. The following prior distributions were used: 
\begin{align*}
    \lambda - \mu &\sim {\rm Gamma}({\rm shape} = 3, {\rm rate} = 1) \\
    \mu/\lambda & \sim {\rm Beta}(5, 5)
\end{align*}
These priors guarantee that $\lambda > \mu$, implying that the probability of extinction is less than 1. 
We evaluate ABC-SMC-RF and ABC-SMC-DRF with $T=4$ iterations, each with $N_t=5,000$ simulations in the reference table and the random forest in each iteration is constructed from $B = 2,500$ trees.
We implement the normal perturbation kernels $K_t(\lambda-\mu| \lambda^*-\mu^*) = \mathcal{N}\left(\lambda^*-\mu^*, 2\cdot\sigma_{t-1}^{\lambda - \mu}\right)$ and $K_t\left(\mu/\lambda\middle|\mu^*/\lambda^*\right) = \mathcal{N}\left(\mu^*/\lambda^*, 2\cdot\sigma_{t-1}^{\mu/\lambda}\right)$, where $\sigma_{t-1}^{\lambda - \mu}$ and $\sigma_{t-1}^{\mu/\lambda}$ are the empirical variances of the Malthusian and extinction parameters from iteration $t-1$.
This perturbation framework was proven to be efficient in the context of ABC-SMC \citep{beaumont_adaptive_2009}.
The values of $\sigma_{t-1}^{\lambda - \mu}$ and $\sigma_{t-1}^{\mu/\lambda}$ are reported in Table \ref{tab:beaumont_variances}.
Figure \ref{fig:bd_reparametrized} present the joint and marginal posterior distributions from ABC-SMC-RF and ABC-SMC-DRF.
The results from both methods are close to the true posterior distribution.

\begin{table}[htbp]
\centering
\begin{tabular}{|l|c|c|c|c|c|c|}
\hline
\multirow{2}{*}{Variance} 
& \multicolumn{3}{c|}{ABC-SMC-RF iterations} 
& \multicolumn{3}{c|}{ABC-SMC-DRF iterations} \\ 
\cline{2-7}
& $t=2$ & $t=3$ & $t=4$ 
& $t=2$ & $t=3$ & $t=4$ \\ 
\hline\hline
$2\cdot\sigma_{t-1}^{\lambda - \mu}$ 
& 0.1207 & 0.1166 & 0.1514 
& 0.2349 & 0.2785 & 0.2187 \\ 
\hline
$2\cdot\sigma_{t-1}^{\mu/\lambda}$
& 0.0273 & 0.0323 & 0.0343 
& 0.0231 & 0.0234 & 0.0209 \\ 
\hline
\end{tabular}
\caption{Empirical variances in ABC-SMC-RF and ABC-SMC-DRF across different iterations}
\label{tab:beaumont_variances}
\end{table}

\subsubsection{Controlled branching process}

We also investigate the application of ABC-SMC-(D)RF for a discrete-time controlled branching process developed by \cite{gonzalez2022abc}.
Let $Z_n$ be the population size at generation $n=0,1,\dots$, which is defined recursively via
\begin{align*}
    &Z_0 = 1 \\
    &Z_{n+1} = \sum_{j=1}^{\phi_n(Z_n)} X_{n,j}
\end{align*}
where $\phi_n(z)\sim\mathrm{Bin}(\xi(z), \gamma)$ with $\xi(z) = z + \lfloor \log z \rfloor$ represents the number of the $z$ individuals in generation $n$ that will be parents of generation $n+1$, and $X_{n,j}$ is the number of offspring from the $j$th individual at generation $n$.
\cite{gonzalez2022abc} modeled $X_{n,j}$ as i.i.d. with respect to $n$ and $j$, with probabilities $\pmb{p}(\kappa) = \{ p_k = \mathbb{P}(X_{n,j} = k), k =0,1,\dots,\kappa\}$ where $\kappa$ is the maximum offspring capacity per individual.
The data is simulated with $\gamma=0.8$, $\kappa=4$ and $\pmb{p}(\kappa)\sim\mathrm{Bin}(4,0.9)$.
From data statistics $\left\{Z_1,\dots,Z_{10}\right\}$ and $\phi_9(Z_9)$ (e.g., the number of parents of the last generation), we seek to find parameters $\theta=\left\{\gamma,\kappa,\pmb{p}(\kappa)\right\}$ from prior distributions
\begin{align*}
    \gamma & \sim \beta(1,1)\\
    \kappa & \sim \mathcal{U}(2, 15) \\
    \pmb{p}(\kappa) & \sim \mathcal{D}(\kappa +1, \alpha_\kappa = (1, \dots, 1)).
\end{align*}
As in \citep{gonzalez2022abc}, we analyze the results by examining the mean offspring count $m=\sum_{k=1}^{\kappa} k\cdot p_k(\kappa)$ instead of $\pmb{p}(\kappa)$.

\begin{figure}[!h]
    \centering
    \includegraphics[width=16cm]{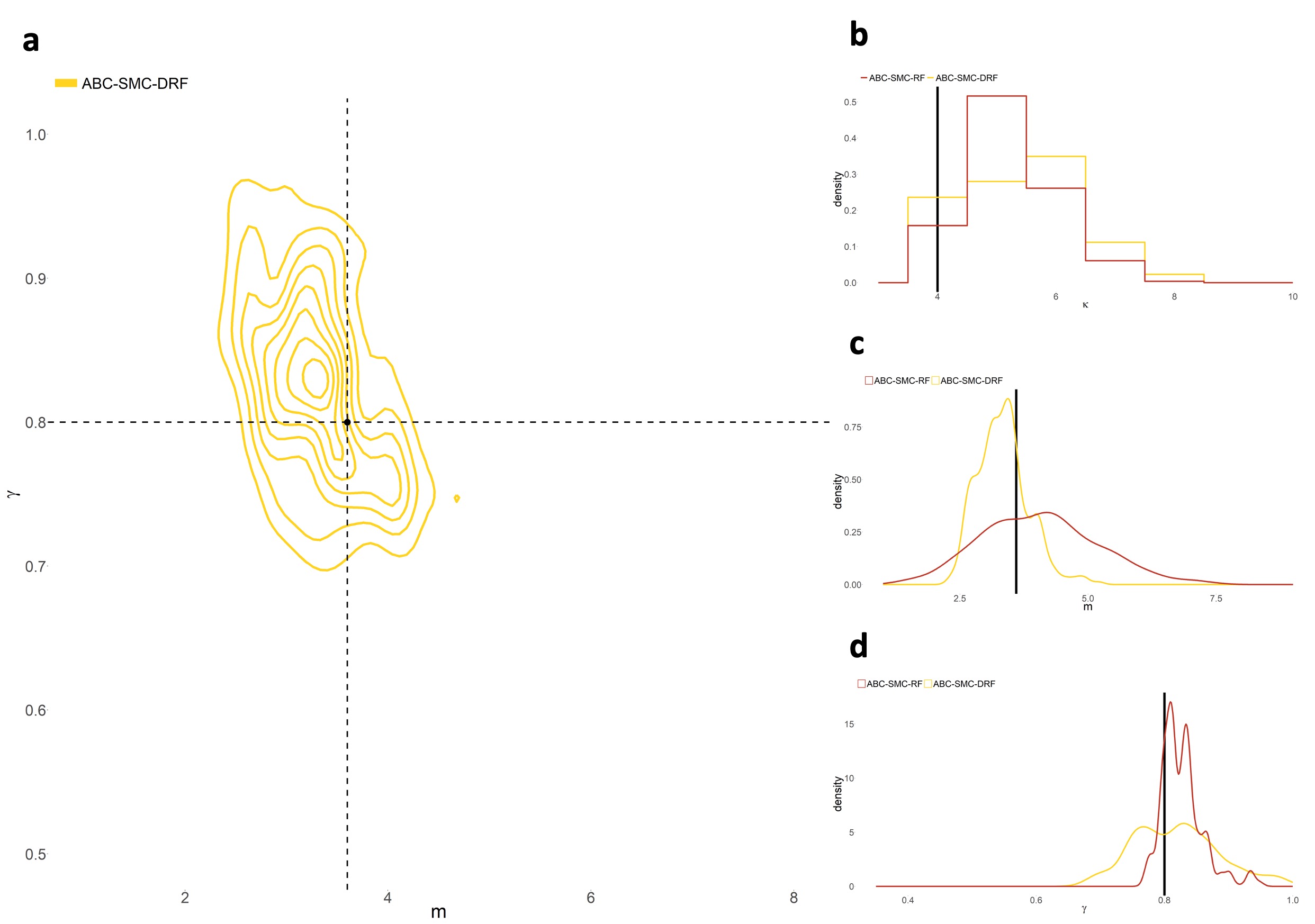}
    \caption{
    Parameter inference for the controlled branching process.
    \textbf{a}:
    Joint posterior distributions for $m$ and $\gamma$ from ABC-SMC-DRF (yellow contours) against true values (broken lines).
    \textbf{b, c, d}:
    Marginal distributions for $\kappa$ (\textbf{b}), $m$ (\textbf{c}) and $\gamma$ (\textbf{d}), inferred with ABC-SMC-RF and ABC-SMC-DRF, compared to true values (black lines).
    }
    \label{fig:CBP}
\end{figure}

Similar to the linear birth-death process above, we perform ABC-SMC-RF and ABC-SMC-DRF with $T=4$ iterations, $N_t=5,000$ simulations per iteration and $B=2,500$ trees in each random forest.
We implement the uniform perturbations $K_t(\kappa|\kappa^*) = \mathcal{U}(\kappa^*-3,\kappa^*+3)$, $K_t(\gamma|\gamma^*) = \mathcal{U}(\gamma^*-0.05,\gamma^*+0.05)$ and $K_t(p_j | p_j^*) = \mathcal{U}(p_j^*-0.05,p_j^*+0.05)$.
Figure \ref{fig:CBP} shows the comparisons between ABC-SMC-(D)RF posterior distributions and the true parameter values $\kappa=4$, $m=3.6$ and $\gamma = 0.8$.

Figure \ref{fig:CBP} shows the inferred marginal posterior distributions of $\kappa$ (Figure \ref{fig:CBP}b), $m$ (Figure \ref{fig:CBP}c), $\gamma$ (Figure \ref{fig:CBP}d), as well as the joint posterior distribution of $(\gamma, m)$ from ABC-SMC-DRF (Figure \ref{fig:CBP}a).
The marginal distributions for $m$ and $\gamma$ from both ABC-SMC-RF and ABC-SMC-DRF are centered around the true values.
However, the ABC-SMC-RF distribution for $m$ is more concentrated compared to ABC-SMC-DRF, and the reverse holds for $\gamma$.
Table \ref{table:cbp_results} also shows that the mean and median of each parameter distribution converge to the true values after successive ABC-SMC-DRF iterations, with decreasing variances.
This results in good approximations for $m$ and $\gamma$, with relative errors $<10\%$ for both parameters in the final iteration.

Both ABC-SMC-RF and ABC-SMC-DRF overestimate $\kappa$ (Figure \ref{fig:CBP}b). 
However, these distributions are similar to the results reported by \cite{gonzalez2022abc} (Figure 1 in \citep{gonzalez2022abc}).
As noted by the authors, the model with maximum offspring count $\kappa$ behaves identically to one with $\kappa'>\kappa$ where $p_{\kappa+1}=\dots=p_{\kappa'}=0$. 
Therefore, $\kappa$ can be inferred to be greater than the true value $\bar\kappa$, as long as the probabilities $p_j$ with $j>\bar\kappa$ are negligible (e.g., $<10^{-3}$).
We note that \citep{gonzalez2022abc} utilizes a multi-step inference framework, where $\kappa$ is estimated first with ABC-SMC and $(\gamma,m)$ are then inferred with ABC-rejection, conditioned on the selected $\kappa$.
In comparison, our framework of using ABC-SMC-(D)RF to infer $\kappa$, $m$ and $\gamma$ simultaneously may be more practical for implementation.

There are multiple sources of variability in the ABC-SMC-(D)RF results, including both the stochastic simulations that form the reference table in each iteration and the construction of the random forests.
Therefore, we perform the ABC-SMC-(D)RF inference for the controlled branching process 500 times and analyze the changes in the inferred distributions (Table \ref{tab:posterior_median_variability}).
Overall, there is some evidence that ABC-SMC-DRF underestimates $m$ and overestimates $\gamma$; the true values of both $m=3.6$ and $\gamma=0.8$ fall outside of their respective 95\% interquantile ranges, $[q_{2.5\%}, q_{97.5\%}]$.
In comparison, the interquantile ranges from ABC-SMC-RF successfully contain the true values, albeit with higher variance for $m$.
This agrees with the results from the single inference run shown in Figure \ref{fig:CBP}.

    \begin{table}[!h]
    \begin{center}
    \begin{tabular}{|c|c|c|c|c|}
        \hline
        \multirow{2}{*}{Statistics} & \multicolumn{4}{c|}{ABC-SMC-DRF iterations} \\
        \cline{2-5}
        & 1 & 2 & 3 & 4 (Final posterior) \\
        \hline\hline
        $\mathbb{E}(\kappa)$ & 6.326 & 5.478 & 5.239 & 5.406 \\
        \hline
        $q_{50\%}(\kappa)$ & 6 & 5 & 5 & 5 \\
        \hline
        Var($\kappa$) & 3.297 & 1.900 & 1.332 & 1.076 \\
        \hline
        $\mathbb{E}(\gamma)$ & 0.833 & 0.856 & 0.854 & 0.823 \\
        \hline
        $q_{50\%}(\gamma)$ & 0.873 & 0.872 & 0.869 & 0.825 \\
        \hline
        Var($\gamma$) & 0.013 & 0.007 & 0.006 & 0.005 \\
        \hline
        $\mathbb{E}(m)$ & 3.368 & 3.225 & 3.258 & 3.353 \\
        \hline
        $q_{50\%}(m)$ & 3.106 & 3.137 & 3.165 & 3.332 \\
        \hline
        Var($m$) & 0.500 & 0.243 & 0.234 & 0.259 \\
        \hline
    \end{tabular}
    \caption{Means, medians and variances of the marginal posterior distributions for the controlled branching process from ABC-SMC-DRF.}
    \label{table:cbp_results}
    \end{center}
\end{table}

\begin{table}[!h]
\centering
\begin{tabular}{|l|c|c|c|c|c|c|c|c|c|}
\hline
\multirow{2}{*}{Method} 
& \multicolumn{3}{|c|}{\(\kappa\)} 
& \multicolumn{3}{|c|}{\(\gamma\)} 
& \multicolumn{3}{|c|}{\(m\)} \\
\cline{2-10}
& $\mathbb{E}$ & $[q_{2.5\%}, q_{97.5\%}]$ & std 
& $\mathbb{E}$ & $[q_{2.5\%}, q_{97.5\%}]$ & std 
& $\mathbb{E}$ & $[q_{2.5\%}, q_{97.5\%}]$ & std \\
\hline\hline
ABC-SMC-RF    &  5.8   &  [5, 7]     &  0.6   
              &  0.817 &  [0.795, 0.841] &  0.012   
              &  4.03  &  [3.05, 5.05]   &  0.53  \\
              \hline
ABC-SMC-DRF   &  5.3   &  [5, 6]     &  0.5   
              &  0.864 &  [0.832, 0.902] &  0.020   
              &  3.34  &  [3.15, 3.56]   &  0.10  \\
\hline
\end{tabular}
\caption{
    Mean ($\mathbb{E}$), $2.5^{th}$ and $97.5^{th}$ percentiles ($[q_{2.5\%}, q_{97.5\%}]$), and standard deviation (std) of ABC-SMC-(D)RF posterior marginal distribution medians for the controlled branching process across 500 runs.
    The data statistics are the same across different inference runs.
}
\label{tab:posterior_median_variability}
\end{table}

\subsubsection{A logistic branching process}\label{logisticbpsect}
Our final branching example concerns the logistic branching process discussed by Lambert \citep{Lambert2005}. This has state space $\{0,1,2,\ldots\}$ and non-zero transition rates given by 
\begin{eqnarray}
 n \to n+1 & \mathrm{at\ rate\ } n \lambda; \nonumber\\
 n \to n-1 & \mathrm{at\ rate\ } n ( \mu + (n-1)c ), \label{allbp}
\end{eqnarray}
for non-negative parameters $\lambda, \mu, c.$ The case $c = 0$ corresponds to the model discussed in Section~\ref{lbpsect}, and we assume henceforth that $c > 0$. Lambert showed that there are two dichotomous behaviors in this model, depending on whether $\mu = 0$ or $\mu > 0$. When $\mu = 0$ the population size $Z(\cdot)$ remains positive and converges in distribution to a Poisson random variable $P$ with mean $\lambda/c$, conditional on $P > 0$. When $\mu > 0$ the population goes extinct with probability 1. If we observe the population size at time points $t_1 < t_2 < \ldots < t_n$ and it happens that $Z(t_n) > 0$, we might ask whether there is evidence that $\mu = 0$ or $\mu > 0$.

This question can be addressed in principle using \texttt{abcrf} and \texttt{drf} approaches, by assuming priors for $\mu$ and $c$ (and, w.l.o.g., $\lambda = 1$), generating a reference set and proceeding as earlier. One object of interest is the posterior probability that $\mu = 0$, which gives information about the likelihood the population will survive.

Rather than detail the results here, we note that this example seems to be a case where the SMC version of the random forest methods is not feasible. 
This is because the algorithms successively sample new particles from the previous iteration and perturb them with Markov kernels, assuming smooth transitions.
The perturbed particles in iterations $t\ge 2$ therefore include $\mu>0$ with probability one, hence ABC-SMC-(D)RF is incapable of accurately evaluating the probability that $\mu=0$.

\section{Discussion}\label{sec:conclusions}

In this paper, we introduce a new Bayesian inference method, Approximate Bayesian Computation sequential Monte Carlo with random forests (ABC-SMC-(D)RF).
It inherits random forest's non-parametric nature, resulting in less dependence on user-defined arguments compared to traditional ABC methods such as ABC-REJ, MCMC and ABC-SMC.

The random forest is embedded in a sequential Monte Carlo regime that progressively updates the parameter distributions to focus on regions in the parameter space with higher likelihood.
We perform numerical experiments for deterministic and stochastic models in ecology, population genetics and systems biology, and observe that ABC-SMC-(D)RF typically results in better posterior approximations than previous RF methods. The Lotka-Volterra example illustrates that generic ABC-SMC-RF produces results comparable  to traditional ABC implementations with optimized arguments.

However, there are several areas of improvement that can extend ABC-SMC-(D)RF's capabilities.
First, a hyperparameter in ABC-SMC-(D)RF that potentially has an impact on the results is the choice of perturbation kernels $K_t\left(\theta\middle|\theta'\right)$, which should balance between exploring the parameter space and targeting the regions already found to contain high likelihood from previous iterations.
In ABC-MCMC and ABC-SMC, it is common to choose uniform or Gaussian kernels, but the optimal kernel form and parameterization for specific problems may be complicated \citep{drovandi2011estimation, lee2012choice}.
A typical approach is to employ kernels that are adjusted dynamically depending on the performance of the previous iteration \citep{beaumont_adaptive_2009, del2012adaptive, liu2000multiple, filippi2013optimality, atchade2010limit}.
By default, ABC-SMC-(D)RF implements Gaussian kernels $K_t\left(\theta\middle|\theta'\right)=\text{Normal}\left(\theta',2\cdot\text{var}\left(\theta_{t-1}\right)\right)$, proposed by \cite{beaumont_adaptive_2009} for ABC-SMC.
Further work might evaluate other adaptive kernel frameworks.

Second, it is sometimes not necessary to continue iteration in ABC-SMC-(D)RF, if further computation is unlikely to yield significant improvements in the posterior distributions.
There have been a number of papers  determining the stopping criterion in ABC-SMC (for example, \cite{prangle2017adapting}), and their incorporation in ABC-SMC-(D)RF promises to lower the computational cost while retaining accuracy.

Finally, in this paper we have focused on parameter inference.
ABC-RF can perform model selection \citep{pudlo2016reliable}, and the same framework could potentially be implemented in ABC-DRF and ABC-SMC-(D)RF.
This will be relevant for problems similar to the logistic branching process in Section \ref{logisticbpsect}, where there is interest in  model selection. ABC-SMC-(D)RF is designed as a wrapper around \texttt{abcrf} \citep{abcrf-packages} and \texttt{drf} \citep{drf-packages} and can therefore be updated together with the original libraries.

\section*{Code availability}
The code for ABC-SMC-(D)RF and studies performed in this paper are available at 
\url{https://github.com/dinhngockhanh/abcsmcrf}.

\section*{Acknowledgments}
The authors acknowledge the support from the Herbert and Florence Irving Institute for Cancer Dynamics at Columbia University.

\section*{Additional information}
The authors declare no competing interests.

\bibliography{References}

\begin{thebibliography}{42}
\providecommand{\natexlab}[1]{#1}
\providecommand{\url}[1]{\texttt{#1}}
\expandafter\ifx\csname urlstyle\endcsname\relax
  \providecommand{\doi}[1]{doi: #1}\else
  \providecommand{\doi}{doi: \begingroup \urlstyle{rm}\Url}\fi

\bibitem[Atchad{\'e} and Fort(2010)]{atchade2010limit}
Y.~Atchad{\'e} and G.~Fort.
\newblock Limit theorems for some adaptive {MCMC} algorithms with subgeometric
  kernels.
\newblock \emph{Bernoulli}, 16:\penalty0 116--154, 2010.

\bibitem[Beaumont et~al.(2002)Beaumont, Zhang, and Balding]{Beaumont2002}
M.~Beaumont, W.~Zhang, and D.~Balding.
\newblock Approximate {B}ayesian computation in population genetics.
\newblock \emph{Genetics}, 162:\penalty0 2025--2035, 2002.

\bibitem[Beaumont et~al.(2009)Beaumont, Cornuet, Marin, and
  Robert]{beaumont_adaptive_2009}
M.~A. Beaumont, J.-M. Cornuet, J.-M. Marin, and C.~P. Robert.
\newblock Adaptive approximate {Bayesian} computation.
\newblock \emph{Biometrika}, 96:\penalty0 983--990, 2009.

\bibitem[Breiman(2001)]{breiman2001random}
L.~Breiman.
\newblock Random forests.
\newblock \emph{Machine Learning}, 45:\penalty0 5--32, 2001.

\bibitem[\'Cevid et~al.(2022)\'Cevid, Michel, Näf, Bühlmann, and
  Meinshausen]{cevid_distributional_2022}
D.~\'Cevid, L.~Michel, J.~Näf, P.~Bühlmann, and N.~Meinshausen.
\newblock Distributional random forests: Heterogeneity adjustment and
  multivariate distributional regression.
\newblock \emph{Journal of Machine Learning Research}, 23:\penalty0 1--79,
  2022.

\bibitem[Degasperi and Gilmore(2008)]{degasperi2007sensitivity}
A.~Degasperi and S.~Gilmore.
\newblock Sensitivity analysis of stochastic models of bistable biochemical
  reactions.
\newblock In M.~Bernardo, P.~Degano, and G.~Zavattaro, editors, \emph{Formal
  Methods for Computational Systems Biology}, volume 5016, pages 1--20.
  Springer-Verlag, Berlin, Heidelberg, 2008.

\bibitem[Del~Moral et~al.(2012)Del~Moral, Doucet, and Jasra]{del2012adaptive}
P.~Del~Moral, A.~Doucet, and A.~Jasra.
\newblock {An adaptive sequential Monte Carlo method for approximate Bayesian
  computation}.
\newblock \emph{Statistics and Computing}, 22:\penalty0 1009--1020, 2012.

\bibitem[Desai and Ouarda(2021)]{desai2021regional}
S.~Desai and T.~B. Ouarda.
\newblock Regional hydrological frequency analysis at ungauged sites with
  random forest regression.
\newblock \emph{Journal of Hydrology}, 594:\penalty0 125861, 2021.

\bibitem[Dinh et~al.(2024)Dinh, Tavar\'e, and Zhang]{dtz24}
K.~N. Dinh, S.~Tavar\'e, and Z.~Zhang.
\newblock Irving institute for cancer dynamics, 2024.
\newblock URL
  \url{https://cancerdynamics.columbia.edu/news/approximate-bayesian-computation-and-distributional-random-forests}.
\newblock Accessed on February 26, 2024.

\bibitem[Drovandi and Pettitt(2011)]{drovandi2011estimation}
C.~C. Drovandi and A.~N. Pettitt.
\newblock Estimation of parameters for macroparasite population evolution using
  approximate {B}ayesian computation.
\newblock \emph{Biometrics}, 67:\penalty0 225--233, 2011.

\bibitem[Filippi et~al.(2013)Filippi, Barnes, Cornebise, and
  Stumpf]{filippi2013optimality}
S.~Filippi, C.~P. Barnes, J.~Cornebise, and M.~P. Stumpf.
\newblock {On optimality of kernels for approximate Bayesian computation using
  sequential Monte Carlo}.
\newblock \emph{Statistical Applications in Genetics and Molecular Biology},
  12:\penalty0 87--107, 2013.

\bibitem[Fu and Li(1997)]{FuLi1997}
Y.-X. Fu and W.-H. Li.
\newblock Estimating the age of the common ancestor of a sample of {DNA}
  sequences.
\newblock \emph{Molecular Biology and Evolution}, 14:\penalty0 195--199, 1997.

\bibitem[Gillespie(1977)]{gillespie1977exact}
D.~T. Gillespie.
\newblock Exact stochastic simulation of coupled chemical reactions.
\newblock \emph{The Journal of Physical Chemistry}, 81:\penalty0 2340--2361,
  1977.

\bibitem[González et~al.(2022)González, Minuesa, and del
  Puerto]{gonzalez2022abc}
M.~González, C.~Minuesa, and I.~del Puerto.
\newblock Approximate {B}ayesian computation approach on the maximal offspring
  and parameters in controlled branching processes.
\newblock \emph{Revista de la Real Academia de Ciencias Exactas, Físicas y
  Naturales. Serie A. Matemáticas}, 116\penalty0 (1):\penalty0 147, 2022.
\newblock \doi{10.1007/s13398-022-01290-w}.

\bibitem[Gretton et~al.(2007)Gretton, Borgwardt, Rasch, Schölkopf, and
  Smola]{10.7551/mitpress/7503.003.0069}
A.~Gretton, K.~M. Borgwardt, M.~Rasch, B.~Schölkopf, and A.~J. Smola.
\newblock {A kernel method for the two-sample problem}.
\newblock In \emph{{Advances in Neural Information Processing Systems 19:
  Proceedings of the 2006 Conference}}. The MIT Press, 2007.

\bibitem[Iooss et~al.(2024)Iooss, Veiga, Janon, Pujol, with contributions~from
  Baptiste~Broto, Boumhaout, Clouvel, Delage, Amri, Fruth, Gilquin, Guillaume,
  Herin, Idrissi, {Le Gratiet}, Lemaitre, Marrel, Meynaoui, Nelson, Monari,
  Oomen, Rakovec, Ramos, Rochet, Roustant, Sarazin, Song, Staum, Sueur, Touati,
  Verges, and Weber]{pkg-sensitivity}
B.~Iooss, S.~D. Veiga, A.~Janon, G.~Pujol, with contributions~from
  Baptiste~Broto, K.~Boumhaout, L.~Clouvel, T.~Delage, R.~E. Amri, J.~Fruth,
  L.~Gilquin, J.~Guillaume, M.~Herin, M.~I. Idrissi, L.~{Le Gratiet},
  P.~Lemaitre, A.~Marrel, A.~Meynaoui, B.~L. Nelson, F.~Monari, R.~Oomen,
  O.~Rakovec, B.~Ramos, P.~Rochet, O.~Roustant, G.~Sarazin, E.~Song, J.~Staum,
  R.~Sueur, T.~Touati, V.~Verges, and F.~Weber.
\newblock \emph{sensitivity: Global Sensitivity Analysis of Model Outputs and
  Importance Measures}, 2024.
\newblock URL \url{https://CRAN.R-project.org/package=sensitivity}.
\newblock R package version 1.30.0.

\bibitem[Jabot et~al.(2023)Jabot, Faure, Dumoulin, and Albert]{easyabc-package}
F.~Jabot, T.~Faure, N.~Dumoulin, and C.~Albert.
\newblock \emph{EasyABC: Efficient Approximate Bayesian Computation Sampling
  Schemes}, 2023.
\newblock URL \url{https://CRAN.R-project.org/package=EasyABC}.
\newblock R package version 1.5.2.

\bibitem[Jung and Marjoram(2011)]{jung2011choice}
H.~Jung and P.~Marjoram.
\newblock Choice of summary statistic weights in {Approximate Bayesian
  Computation}.
\newblock \emph{Statistical Applications in Genetics and Molecular Biology},
  10:\penalty0 art. 45, 2011.

\bibitem[Keiding(1975)]{nk1975}
N.~Keiding.
\newblock Maximum likelihood estimation in the birth-and-death process.
\newblock \emph{The Annals of Statistics}, 3:\penalty0 363--372, 1975.

\bibitem[Kendall(1948)]{kendall1948generalized}
D.~G. Kendall.
\newblock On the generalized ``birth-and-death" process.
\newblock \emph{The Annals of Mathematical Statistics}, 19:\penalty0 1--15,
  1948.

\bibitem[Lambert(2005)]{Lambert2005}
A.~Lambert.
\newblock The branching process with logistic growth.
\newblock \emph{The Annals of Applied Probability}, 15:\penalty0 1506--1535,
  2005.

\bibitem[Lee(2012)]{lee2012choice}
A.~Lee.
\newblock {On the choice of MCMC kernels for approximate Bayesian computation
  with SMC samplers}.
\newblock In \emph{Proceedings of the 2012 Winter simulation conference (WSC)},
  pages 1--12. IEEE, 2012.

\bibitem[Liu et~al.(2000)Liu, Liang, and Wong]{liu2000multiple}
J.~S. Liu, F.~Liang, and W.~H. Wong.
\newblock The multiple-try method and local optimization in {M}etropolis
  sampling.
\newblock \emph{Journal of the American Statistical Association}, 95:\penalty0
  121--134, 2000.

\bibitem[Lotka(1925)]{Lotka_elements_1925}
A.~J. Lotka.
\newblock \emph{Elements of Physical Biology}.
\newblock Williams and Wilkins Co., London, 1925.

\bibitem[Marin et~al.(2018)Marin, Pudlo, Estoup, and
  Robert]{marin2018likelihood}
J.-M. Marin, P.~Pudlo, A.~Estoup, and C.~Robert.
\newblock Likelihood-free model choice.
\newblock In S.~A. Sisson, Y.~Fan, and M.~Beaumont, editors, \emph{Handbook of
  Approximate Bayesian Computation}, pages 153--178. Chapman and Hall/CRC,
  2018.

\bibitem[Marin et~al.(2022)Marin, Raynal, Pudlo, Robert, and
  Estoup]{abcrf-packages}
J.-M. Marin, L.~Raynal, P.~Pudlo, C.~P. Robert, and A.~Estoup.
\newblock \emph{abcrf: Approximate Bayesian Computation via Random Forests},
  2022.
\newblock URL \url{https://CRAN.R-project.org/package=abcrf}.
\newblock R package version 1.9.

\bibitem[Marjoram et~al.(2003)Marjoram, Molitor, Plagnol, and
  Tavar{\'e}]{marjoram2003markov}
P.~Marjoram, J.~Molitor, V.~Plagnol, and S.~Tavar{\'e}.
\newblock Markov chain {M}onte {C}arlo without likelihoods.
\newblock \emph{Proceedings of the National Academy of Sciences}, 100:\penalty0
  15324--15328, 2003.

\bibitem[Michel and \'Cevid(2021)]{drf-packages}
L.~Michel and D.~\'Cevid.
\newblock \emph{drf: Distributional Random Forests}, 2021.
\newblock URL \url{https://CRAN.R-project.org/package=drf}.
\newblock R package version 1.1.0.

\bibitem[Monari and Strachan(2017)]{monari2017characterization}
F.~Monari and P.~Strachan.
\newblock Characterization of an airflow network model by sensitivity analysis:
  parameter screening, fixing, prioritizing and mapping.
\newblock \emph{Journal of Building Performance Simulation}, 10:\penalty0
  17--36, 2017.

\bibitem[Morris(1991)]{morris1991factorial}
M.~D. Morris.
\newblock Factorial sampling plans for preliminary computational experiments.
\newblock \emph{Technometrics}, 33:\penalty0 161--174, 1991.

\bibitem[Prangle(2017)]{prangle2017adapting}
D.~Prangle.
\newblock Adapting the {ABC} distance function.
\newblock \emph{Bayesian Analysis}, 12:\penalty0 289--309, 2017.

\bibitem[Pritchard et~al.(1999)Pritchard, Seielstad, Perez-Lezaun, and
  Feldman]{pritchard1999population}
J.~K. Pritchard, M.~T. Seielstad, A.~Perez-Lezaun, and M.~W. Feldman.
\newblock Population growth of human {Y} chromosomes: a study of {Y} chromosome
  microsatellites.
\newblock \emph{Molecular Biology and Evolution}, 16:\penalty0 1791--1798,
  1999.

\bibitem[Pudlo et~al.(2016)Pudlo, Marin, Estoup, Cornuet, Gautier, and
  Robert]{pudlo2016reliable}
P.~Pudlo, J.-M. Marin, A.~Estoup, J.-M. Cornuet, M.~Gautier, and C.~P. Robert.
\newblock Reliable {ABC} model choice via random forests.
\newblock \emph{Bioinformatics}, 32:\penalty0 859--866, 2016.

\bibitem[Raynal et~al.(2019)Raynal, Marin, Pudlo, Ribatet, Robert, and
  Estoup]{raynal2019abc}
L.~Raynal, J.-M. Marin, P.~Pudlo, M.~Ribatet, C.~P. Robert, and A.~Estoup.
\newblock {ABC} random forests for {B}ayesian parameter inference.
\newblock \emph{Bioinformatics}, 35:\penalty0 1720--1728, 2019.

\bibitem[Rigatti(2017)]{rigatti2017random}
S.~J. Rigatti.
\newblock Random forest.
\newblock \emph{Journal of Insurance Medicine}, 47:\penalty0 31--39, 2017.

\bibitem[Segal(2004)]{segal2004machine}
M.~R. Segal.
\newblock Machine learning benchmarks and random forest regression.
\newblock Technical report, UCSF: Center for Bioinformatics and Molecular
  Biostatistics, 2004.
\newblock URL \url{Retrieved from https://escholarship.org/uc/item/35x3v9t4}.

\bibitem[Sisson et~al.(2018)Sisson, Fan, and Beaumont]{sisson2018handbook}
S.~A. Sisson, Y.~Fan, and M.~Beaumont, editors.
\newblock \emph{Handbook of Approximate Bayesian Computation}.
\newblock CRC Press, 2018.

\bibitem[Tavar{\'e}(2018)]{tavare2018linear}
S.~Tavar{\'e}.
\newblock The linear birth--death process: an inferential retrospective.
\newblock \emph{Advances in Applied Probability}, 50\penalty0 (A):\penalty0
  253--269, 2018.

\bibitem[Tavar{\'e} et~al.(1997)Tavar{\'e}, Balding, Griffiths, and
  Donnelly]{tavare1997inferring}
S.~Tavar{\'e}, D.~J. Balding, R.~C. Griffiths, and P.~Donnelly.
\newblock Inferring coalescence times from {DNA} sequence data.
\newblock \emph{Genetics}, 145:\penalty0 505--518, 1997.

\bibitem[Toni et~al.(2008)Toni, Welch, Strelkowa, Ipsen, and
  Stumpf]{toni_approximate_2008}
T.~Toni, D.~Welch, N.~Strelkowa, A.~Ipsen, and M.~P. Stumpf.
\newblock Approximate {Bayesian} computation scheme for parameter inference and
  model selection in dynamical systems.
\newblock \emph{Journal of The Royal Society Interface}, 6:\penalty0 187--202,
  2008.

\bibitem[Volterra(1928)]{volterra_variations_1928}
V.~Volterra.
\newblock Variations and fluctuations of the number of individuals in animal
  species living together.
\newblock \emph{ICES Journal of Marine Science}, 3:\penalty0 3--51, 1928.

\bibitem[Wilkinson(2018)]{wilkinson2018stochastic}
D.~J. Wilkinson.
\newblock \emph{{Stochastic Modelling for Systems Biology}}.
\newblock Chapman and Hall/CRC, 2018.

\end{thebibliography}
\bibliographystyle{abbrvnat}

\appendix
\section*{Supplementary Materials}
\section{Dependence of ABC-SMC-(D)RF on the prior distributions}

We examine the influence of the weight recalibration in ABC-SMC-(D)RF (Step 15, Algorithm \ref{alg:ABC SMC-RF multi} and Step 18, Algorithm \ref{alg:ABC SMC-RF single}) on the posterior distributions, by comparing ABC-DRF and ABC-SMC-DRF for the hierarchical Normal mean model in Section \ref{sec-hierarchical}.
This example presents a case study for whether ABC-SMC-DRF converges to the correct posterior distribution when the prior distribution is highly non-uniform.
The model, statistics and true joint posterior distribution are given in Section \label{sec-hierarchical}.

\begin{figure}[H]
    \centering
    \includegraphics[width=14cm]{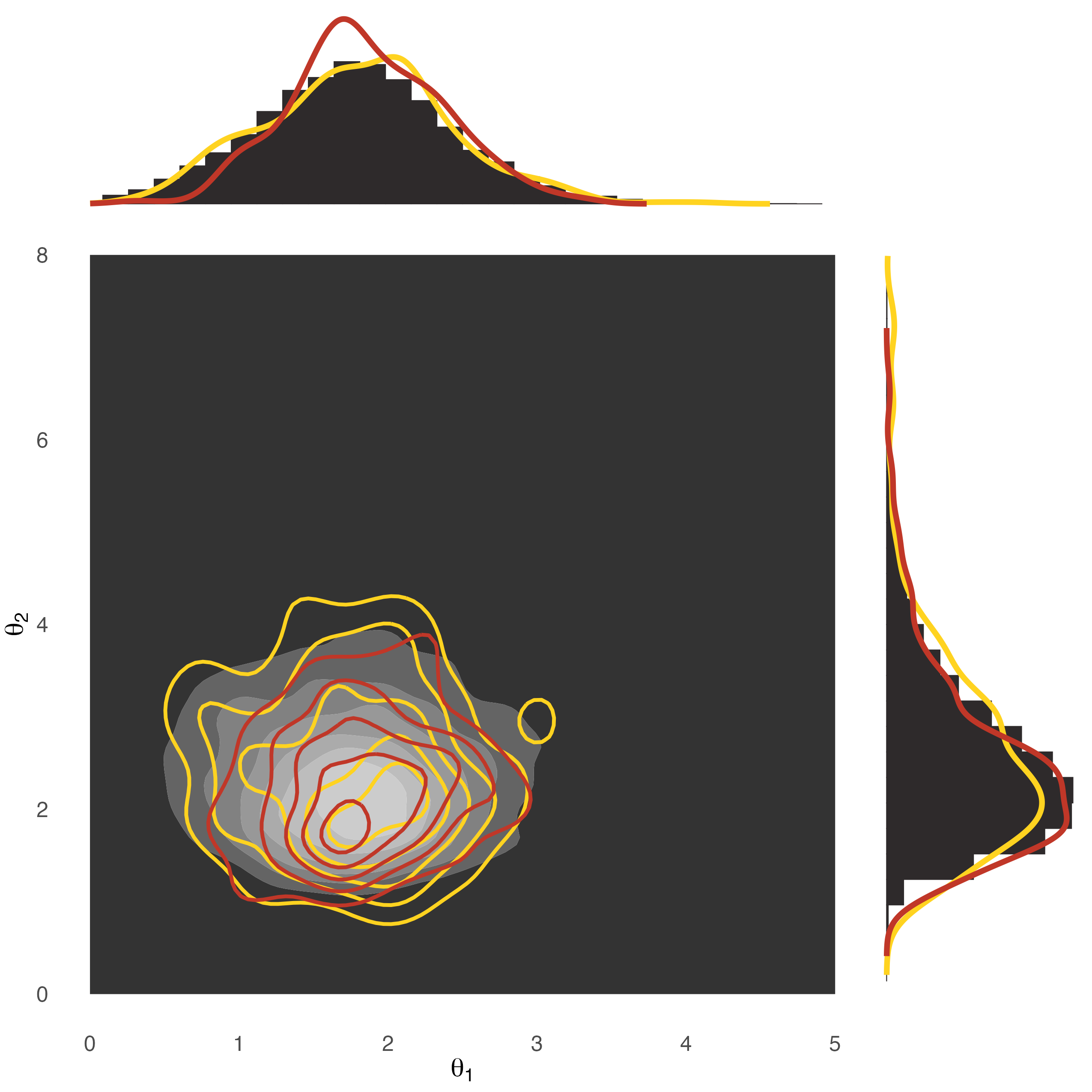}
    \caption{
    Inference of $\theta=(\theta_1,\theta_2)$ in the hierarchical model, with $\alpha=4,\beta=5$.
    Joint posterior distributions for $\theta_1$ and $\theta_2$,
    inferred from ABC-DRF with $N=20,000$ simulations (yellow contours)
    and ABC-SMC-DRF with $N_t=20,000$ simulations for iterations $t=1,\dots,20$ (red contours)
    against ground truth (density heatmap in gray-scale), with marginal distributions for each parameter from ABC-DRF (yellow histogram) and ABC-SMC-DRF (red histogram) against ground truth (black histogram).
    }
    \label{fig:toy}
\end{figure}

In Fig. \ref{fig:toy}, we compare the results from \texttt{drf} with 20,000 particles, and \texttt{abcsmcrf} with 20,000 particles per iteration and 20 iterations in total using Gaussian perturbations with adaptive variances formulated in \citep{beaumont_adaptive_2009}.
As discussed in the main text, ABC-DRF can already approximate the true posterior distribution well.
Successive ABC-SMC-DRF posteriors essentially remain constant and in agreement with the true distribution.
This demonstrates that the weight calibrations in ABC-SMC-(D)RF, adopted from ABC-SMC, maintain the influence of the prior distributions and particle perturbations on the posterior distributions, and that ABC-SMC-(D)RF becomes stable as the iterations proceed. 

\end{document}